\documentclass[journal,twocolumn,10pt]{IEEEtran}
\usepackage{graphicx}
\usepackage{amssymb}
\usepackage{mathtools}
\usepackage{dsfont}
\usepackage{cite}
\usepackage{stfloats}
\usepackage{subfigure}
\usepackage{psfrag}
\usepackage[mathscr]{euscript}
\usepackage{acronym}  % make an acronym
\usepackage{amsmath}
\usepackage{algorithm}
\usepackage[noend]{algpseudocode}
\usepackage{booktabs}
\usepackage{bbm}

\makeatletter
\def\BState{\State\hskip-\ALG@thistlm}
\makeatother

\usepackage{float}
\usepackage{balance}

\acrodef{CCDF}{complementary cumulative distribution function}
\acrodef{CF}{characteristic function}
\acrodef{PPP}{Poisson point processe}
\acrodef{RV}{random variable}
\acrodef{i.i.d.}{independent and identically distributed}
\acrodef{PDF}{probability distribution function}
\acrodef{CDF}{cumulative distribution function}
\acrodef{ch.f.}{characteristic function}
\acrodef{AWGN}{additive white Gaussian noise}
\acrodef{SNR}{signal-to-noise ratio}
\acrodef{LRT}{likelihood ratio test}
\acrodef{DRT}{distance ratio test}
\acrodef{GLRT}{generalized likelihood ratio test}
\acrodef{CRLB}{Cram\'{e}r-Rao lower bound}
\acrodef{CRB}{Cram\'{e}r-Rao bound}
\acrodef{ZZLB}{Ziv-Zakai lower bound}
\acrodef{ZZB}{Ziv-Zakai bound}
\acrodef{LOS}{line-of-sight}
\acrodef{ToF}{time-of-flight}
\acrodef{NLOS}{non-line-of-sight}
\acrodef{GDOP}{geometric dilution of precision}
\acrodef{GPS}{Global Positioning System}
\acrodef{FIM}{Fisher information matrix}
\acrodef{PEB}{position error bound}
\acrodef{SPEB}{squared position error bound}
\acrodef{TOA}{time-of-arrival}
\acrodef{TOF}{time-of-flight}
\acrodef{WSN}{wireless sensor network}
\acrodef{MAC}{medium access control}
\acrodef{RSS}{received signal strength}
\acrodef{WAF}{wall attenuation factor}
\acrodef{TDOA}{time difference-of-arrival}
\acrodef{RF}{radiofrequency}
\acrodef{RTT}{round-trip time}
\acrodef{AOA}{angle-of-arrival}
\acrodef{MF}{matched filter}
\acrodef{ED}{energy detector}
\acrodef{ML}{maximum likelihood}
\acrodef{MSE}{mean-square error}
\acrodef{RMSE}{root-mean-square error}
\acrodef{LEO}{localization error outage}
\acrodef{ppm}{part-per-million}
\acrodef{ACK}{acknowledge}
\acrodef{UWB}{Ultrawide bandwidth}
\acrodef{TNR}{threshold-to-noise ratio}
\acrodef{LS}{least squares}
\acrodef{IR-UWB}{impulse radio UWB}
\acrodef{FCC}{Federal Communications Commission}
\acrodef{TH}{time-hopping}
\acrodef{PPM}{pulse position modulation}
\acrodef{MUI}{multi-user interference}
\acrodef{PDP}{power delay profile}
\acrodef{BPZF}{band-pass zonal filter}
\acrodef{SIR}{signal-to-interference ratio}
\acrodef{SINR}{signal-to-interference-plus-noise ratio}
\acrodef{RFID}{radio frequency identification}
\acrodef{WPAN}{wireless personal area network}
\acrodef{WWB}{Weiss-Weinstein bound}
\acrodef{DP}{direct path}
\acrodef{MF}{matched filter}
\acrodef{MMSE}{minimum-mean-square-error}
\acrodef{SBS}{serial backward search}
\acrodef{SBSMC}{serial backward search for multiple clusters}
\acrodef{NBI}{narrowband interference}
\acrodef{WBI}{wideband interference}
\acrodef{INR}{interference-to-noise ratio}
\acrodef{CR}{channel response}
\acrodef{CIR}{channel impulse response}
\acrodef{CR}{channel  response}
\acrodef{RADAR}{radar}
\acrodef{MUR}{Multistatic radar}
\acrodef{JBSF}{jump back and search forward}
\acrodef{HDSA}{high-definition situation-aware}
\acrodef{RRC}{root raised cosine}
\acrodef{ST}{simple thresholding}
\acrodef{BTB}{Bellini-Tartara bound}
\acrodef{P-Max}{$P$-Max}  %suggestion, use with \acl{P-Max}
\acrodef{MIMO}{multiple-input multiple-output}
\acrodef{MAP}{maximum a posteriori}
\acrodef{FG}{factor graph}
\acrodef{OP}{outage probability}
\acrodef{WED}{wall extra delay}
\acrodef{RMS}{root mean square}
\acrodef{SPAWN}{sum-product algorithm over a wireless network}
\acrodef{MDD}{minimum distance distribution}
\acrodef{MAP}{maximum a posteriori probability}
\acrodef{SAP}{small cell access point}
\acrodef{UE}{user equipment}
\acrodef{MBS}{macro cell base station}
\acrodef{UER}{\ac{UE} Relay}
\acrodef{D2D}{device-to-device}
\acrodef{MBS}{macro base station}
\acrodef{CSI}{channel state information}
\acrodef{OGR}{outage guard region}
\acrodef{FUR}{feasible UER region}
\acrodef{EHR}{energy harvesting region}
\acrodef{EH}{energy harvesting}
\acrodef{D2D-EHSN}{D2D communication provided \ac{EH} small cell network}
\acrodef{D2D-EHHN}{D2D communication provided \ac{EH} heterogeneous network}
\acrodef{3GPP}{3rd Generation Partnership Project}
\acrodef{BS}{base station}
\acrodef{DF}{decode and forward}
\acrodef{CCDF}{complementary cumulative distribution function}
\acrodef{ZF}{zero forcing}
\acrodef{RZF}{regularized zero forcing}
\acrodef{WLLN}{weak law of large number}
\acrodef{SLLN}{strong law of large numbers}
\acrodef{TDD}{Time-division duplex}
\acrodef{EE}{energy efficiency} 
\acrodef{HetNet}{heterogeneous network} 
\acrodef{SCP}{Single Cell Processing}
\acrodef{CBF}{Coordinated Beamforming}
\usepackage{color}
\usepackage{dsfont}
\usepackage{bbm}

\DeclareMathAlphabet{\mathsf}{OML}{cmbr}{m}{it}

\newtheorem{theorem}{\bf Theorem}
\newtheorem{lemma}{\bf Lemma}
\newtheorem{corollary}{\bf Corollary}

\newtheorem{remark}{\bf Remark}

\newtheorem{assumption}{\bf Assumption}

\newcommand{\bd}{\begin{description}}
\newcommand{\ed}{\end{description}}
\newcommand{\be}{\begin{enumerate}}
\newcommand{\ee}{\end{enumerate}}
\newcommand{\bi}{\begin{itemize}}
\newcommand{\ei}{\end{itemize}}
\newcommand{\bl}{\begin{list}}
\newcommand{\el}{\end{list}}
\newcommand{\bt}{\begin{tabbing}}
\newcommand{\et}{\end{tabbing}}

\newcommand{\paperTitle}{ Understanding Age of Information in Large-Scale Wireless Networks }

\begin{document}

{
\title{\paperTitle}

\author{

	    Howard~H.~Yang, \textit{Member, IEEE},
        Chao Xu, \textit{Member, IEEE},
        Xijun Wang, \textit{Member, IEEE},\\
        Daquan Feng, \textit{Member, IEEE},
	    and Tony~Q.~S.~Quek, \textit{Fellow, IEEE}
     %  \textit{Singapore University of Technology and Design, Singapore}

% \thanks{Manuscript received Feb. 03, 2018, revised Mar. 31, and May 28, 2018, and accepted May 29, 2018. The associate editor coordinating the review of this letter and
%     approving it for publication was Dr. Chun Tung Chou.

%     This work was supported in part by the MOE ARF Tier 2 under Grant MOE2015-T2-2-104 and in part by the SUTD-ZJU Research Collaboration under Grant SUTD-ZJU/RES/01/2016.}

\thanks{H.~H.~Yang is with the Zhejiang University/University of Illinois at Urbana-Champaign Institute, Zhejiang University, Haining 314400, China (e-mail: haoyang@intl.zju.edu.cn).}
\thanks{T.~Q.~S.~Quek is with the Information System Technology and Design Pillar, Singapore University of Technology and Design, Singapore 487372 (e-mail: tonyquek@sutd.edu.sg).}
\thanks{ C. Xu is with the School of Information Engineering, Northwest A \& F University, Yangling, Shaanxi, China (e-mail: cxu@nwafu.edu.cn).  }
\thanks{ X. Wang is with the School of Electronics and Communication Engineering, Sun Yat-sen University, Guangzhou, China (e-mail: wangxijun@mail.sysu.edu.cn). }
\thanks{ D. Feng is with the College of Information Engineering, Shenzhen University, Shenzhen 518060, China (e-mail: fdquan@szu.edu.cn). }
% \thanks{Y.~Zhong is with the School of Electronic Information and Communications, Huazhong University of Science and Technology, Wuhan, P.R. China (email: yzhong@hust.edu.cn).}
% \thanks{H.~H.~Yang and T.~Q.~S.~Quek are with the Singapore University of Technology and Design (e-mail: howard\_yang@sutd.edu.sg, tonyquek@sutd.edu.sg). Y.~Wang is with Nanjing University of Post and Telecommunications (e-mail: wangy1585@163.com). }
}
\maketitle
\acresetall
\thispagestyle{empty}
\begin{abstract}
The notion of age-of-information (AoI) is investigated in the context of large-scale wireless networks, in which transmitters need to send a sequence of information packets, which are generated as independent Bernoulli processes, to their intended receivers over a shared spectrum. Due to interference, the rate of packet depletion at any given node is entangled with both the spatial configurations, which determine the path loss, and temporal dynamics, which influence the active states, of the other transmitters, resulting in the queues to interact with each other in both space and time over the entire network.
To that end, variants in the packet update frequency affect not just the inter-arrival time but also the departure process, and the impact of such phenomena on the AoI is not well understood.
In this paper, we establish a theoretical framework to characterize the AoI performance in the aforementioned setting. Particularly, tractable expressions are derived for both the peak and average AoI under two different transmission protocols, namely the first-come-first-serve (FCFS) and the last-come-first-serve with preemption (LCFS-PR). Additionally, our analysis also accounts for the effects of channel access controls such as ALOHA on the AoI.
The accuracy of the analysis is verified via simulations, and based on the theoretical outcomes, we find that: $i$) networks operating under LCFS-PR are able to attain smaller values of peak and average AoI than that under FCFS, whereas the gain is more pronounced when the infrastructure is densely deployed, $ii$) in sparsely deployed networks, ALOHA with a universally designed channel access probability is not instrumental in reducing the AoI, thus calling for more advanced channel access approaches, and $iii$) when the infrastructure is densely rolled out, there exists a non-trivial ALOHA channel access probability that minimizes the peak and average AoI under both FCFS and LCFS-PR.
\end{abstract}

\acresetall

% ============================================ %
%         Section: Introduction                %
% ============================================ %
\section{Introduction}\label{sec:intro}
Fueled by the eagerness for fresh data in many real-time applications, the age-of-information (AoI) has been introduced as a metric that measures the ``freshness'' of information delivered over a period of time \cite{KauYatGru:12}. Armed with such a metric, a host of new schemes have been developed to achieve timely updating by minimizing AoI, and thus ensuring fresh data for various applications \cite{KosPapAng:17}.
Nonetheless, most of the existing studies were conducted in the point-to-point setting and leaving the AoI performance in the context of wireless networks not well understood.
The central thrust of this article is to fill this research gap with special emphasis on the stochastic analysis of AoI over a large-scale wireless network.

\subsection{Background and Motivation}
Compared with the conventionally transmitter-centric metrics such as delay or throughput, the AoI puts the focus on the receiver side and measures the time elapsed since the latest packet has been delivered, thus being able to gauge the “freshness” associated with the information packets.
As a result, since the genesis of the AoI \cite{KauYatGru:12}, this concept has received spiralling attention in the literature, especially for applications that have a stringent requirement for timely information updating, for instance the positioning, command/control, or monitoring sensors, because fresh data can be attained via updating approaches that minimize AoI.
The AoI minimization problem has been broadly investigated in the field of queueing theory, under different queueing models \cite{KauYatGru:12,KamKomEph:14,CosCodEph:16} and transmission protocols, including the first-come-first-serve (FCFS) \cite{KauYatGru:12,HuaMod:15}, last-come-first-serve (LCFS) \cite{CosCodEph:16,Yat:18}, with \cite{YatKau:18} or without preemption \cite{CosCodEph:16}, together with different buffer sizes \cite{CosCodEph:16}.
In the presence of wireless communication links, AoI-aware scheduling policies have been proposed to ensure timely information delivery \cite{KadUysSin:16,KadSinUys:18,TalKarMod:18,YanAraQue:20ICASSP,DevDurFer:19,HeYuaEph:16,talak2018optimizing}.
Specifically, because the qualities of radio channels vary across time, a number of approaches have been proposed that exploit such property to strike a balance between the performance of link throughput and information freshness \cite{KadUysSin:16,KadSinUys:18,TalKarMod:18,YanAraQue:20ICASSP}.
Moreover, recognizing transmissions over the wireless medium can be unreliable, \cite{DevDurFer:19} carried out a study to investigate several stop-feedback coding schemes, e.g., the automatic repetition request (ARQ) or hybrid ARQ (HARQ) on the purpose of enhancing AoI.
Besides, due to the shared nature of the spectrum, concurrent transmissions on the same channel can deteriorate the performance of others' via the interference they caused. In that respect, scheduling policies have been proposed to construct the set of simultaneously active links while pertaining to an acceptable interference level \cite{HeYuaEph:16,talak2018optimizing}.
In the particular context of internet-of-things (IoT), AoI minimizing schemes that prioritize the transmission orders by accounting for transmission and computing \cite{XuYanWan:19} or reducing energy consumption \cite{CorRohGun:19} have also been explored.

However, these results are devised based on abstract models that assume a certain distribution to the packet departure process and do not precisely capture the impact of physical transmission environments. In fact, wireless communications are generally subject to ($a$) the spatial configuration of transceivers because that affects the path loss, and ($b$) the active states of co-channel transmitters which influence the interference.
Moreover, owing to the broadcast nature of the wireless medium, the transmitters sharing a common spectrum in space will interact with each other through the interference they cause.
As a result, the queueing dynamics at any transmitter is entangled with those of its neighboring nodes. The phenomena of such space-time interactions are commonly known as the \textit{spatially interacting queues}.
To understand the impact of this phenomenon on the performance of communication links, a line of recent studies \cite{YanGerZho:17,WanYanZhu:19WCL,ZhoQueGe:16,GhaElsBad:17,YanWanQue:18,YanQue:19,ChiElSCon:17,ChiElSCon:19,HuZhoZha:18} have been carried out, which combined the stochastic geometry with queueing theory to produce closed-form expressions for various network statistics, including the throughput \cite{YanGerZho:17,WanYanZhu:19WCL}, delay \cite{ZhoQueGe:16,GhaElsBad:17,YanWanQue:18}, coverage probability \cite{YanQue:19,ChiElSCon:17,ChiElSCon:19}, and even AoI \cite{HuZhoZha:18}.
Nevertheless, these analysis are derived either based on the \textit{favorable/dominant system argument} \cite{ZhoQueGe:16,HuZhoZha:18}, which often resulted in upper/lower bounds that are too loose to unveil useful information, or rely on the \textit{mean-field approximation} \cite{GhaElsBad:17,YanWanQue:18,YanQue:19,ChiElSCon:17,ChiElSCon:19} that assumes the queues evolve independently from each other and ignores the intrinsic interaction amongst the queues.
In wireless networks that are densely deployed, which is the modern trend of ``scaling up the architecture" \cite{CorRohGun:19}, transmitters in proximity inevitably interact with each other and hence have correlated queues. To this end, the mean-field approximation is no longer a valid assumption and a more sophisticated analysis that accounts for all these effects is of necessity toward a thorough understanding of the AoI performance in large-scale networks so as to facilitate any further designs.
%A theoretical template that accounts for all these effects is essential for the thorough understanding of AoI performance in large-scale networks, so as to facilitate any further designs.

\subsection{Approach and Summary of Results}
In this paper, we leverage the Poisson bipolar network to model the spatial deployment of transmitters and receivers.
The dynamics of status updating at each transmitter is modeled as a discrete-time queueing system, in which the information packets are generated according to independent Bernoulli processes and are stored in an infinite-capacity buffer.
The transmitters initiate channel access attempts for packet transmissions at each time slot if their buffers are non-empty.
And the transmissions are successful only if the received signal-to-interference-plus-noise ratio (SINR) exceeds a decoding threshold, upon which the packet can be removed from the buffer.
In this network, we consider two different protocols, namely the first-come-first-serve (FCFS) and last-come-first-serve with preemption (LCFS-PR), to schedule the packet transmission order.
Additionally, we adopt ALOHA to control the interference level by approving channel access attempts from each node with a fixed probability.
By jointly using tools from stochastic geometry and queueing theory, we derive accurate and tractable expressions for both the peak and average AoI.
The analytical results enable us to explore the effect from various network parameters on the AoI performance and hence devise useful insights for further design options.
Our main contributions are summarized below.
\begin{itemize}
\item We develop a theoretical template for the understanding of AoI in large-scale wireless networks. The proposed framework is general and encompasses all the key features of a wireless system, including not just the channel gain and interference, but more importantly, the interplay between the geographic locations of information sources and the dynamics of their status updating.
\item We derive an accurate expression for the transmission success probability, which facilitates the quantitative assessment of conditions for the queues to remain stable in large-scale wireless networks as well as the AoI under both FCFS and LCFS-PR protocols. Different from \cite{GhaElsBad:17,YanWanQue:18,YanQue:19,ChiElSCon:17,ChiElSCon:19}, our analysis does not rely on the mean-field approximation in space and directly tackles the crux of spatially interacting queues by extracting a non-homogeneous PPP from the homogeneous setup to model the locations of interferers.
\item Our analysis reveals that: $i$) networks operating under LCFS-PR is able to attain smaller values of peak and average AoI than that under FCFS, whereas the gain is more pronounced when the infrastructure is densely deployed, $ii$) in sparsely deployed networks, simple channel access control such as ALOHA is not instrumental to reduce the AoI, thus calling for more advanced approaches, and $iii$) when the infrastructure is densely rolled out, there exists a non-trivial ALOHA channel access probability that minimizes the peak and average AoI under both FCFS and LCFS-PR.
\end{itemize}

The remainder of the paper is organized as follows. We introduce the system model in Section II. In Section III, we detail the analysis of transmission success probability as well as the derivation of the average and peak AoI. We show the simulation and numerical results in Section IV, that confirm the accuracy of our analysis and provide insights about the AoI performance of a large-scale wireless network. We conclude the paper in Section V.

%%%%%%%%%%%%%%%%%%%%%%%%%%%%%%%%%%%%%%%%%%%%%%%%%%%%
\begin{table}
\caption{Notation Summary
%\mynote{revise notation}
} \label{table:notation}
\begin{center}
%\rowcolors{2}%{green!10!yellow}{}
%{cyan!15!}{}
\renewcommand{\arraystretch}{1.3}
%\begin{tabular}{c  p{6.0cm} }
\begin{tabular}{c  p{5.5cm} }
\hline
{\bf Notation} & {\hspace{2.5cm}}{\bf Definition}
\\
%\midrule
\hline
$\tilde{\Phi}$; $\lambda$ & PPP modeling the locations of transmitters; transmitter spatial deployment density \\
$\bar{\Phi}$; $\lambda$ & PPP modeling the locations of receivers; receiver spatial deployment density \\
$\Phi$ & Superposition of the PPPs $\tilde{\Phi}$ and $\hat{\Phi}$, i.e., $\Phi = \tilde{\Phi} \cup \hat{\Phi}$ \\
$P_{\mathrm{tx}}$; $\alpha$ &  Transmit power; path loss exponent \\
$\xi$; $\theta$ & Packet update frequency; SINR decoding threshold \\
$r$; $p$ & Distance between a pair of transmitter-receiver nodes; channel access probability for each transmitter with non-empty buffer \\
$\mu_{0, t}^\Phi$ & Transmission success probability of link $0$ at time slot $t$, conditioned on the point process $\Phi$ \\
$a_j^\Phi$ & Queue non-empty probability at transmitter node $j$, conditioned on the point process $\Phi$ \\
$A^{\mathrm{ave}}$; $A^{\mathrm p}$ & Network average AoI; network peak AoI \\
\hline
\end{tabular}
\end{center}\vspace{-0.63cm}
\end{table}%

% ============================================ %
%         Section: System Model                %
% ============================================ %
\section{System Model}\label{sec:sysmod}
In this section, we detail the configuration of our network model, as well as the concepts of average and peak AoI in a large-scale wireless system. The main notations used throughout the paper are summarized in Table~\ref{table:notation}.

\subsection{Network Structure}
As illustrated in Figure~\ref{fig:NetMod_V1}, we setup the spatial configuration of the network using the Poisson bipolar model\footnote{ This type of point process is the epitome of spatial models for infrastructureless networks, e.g., the ad-hoc, D2D, or IoT network \cite{BacBla:09}. }, which consists of a set of transmitters and their corresponding receivers, all located in the Euclidean plane. The transmitting nodes are scattered according to a homogeneous Poisson point process (PPP) $\tilde{\Phi}$ of spatial density $\lambda$. Each transmitter located at $X_i \in \tilde{\Phi}$ has a dedicated receiver, whose location $y_i$ is at distance $r$ in a random orientation.
According to the displacement theorem \cite{BacBla:09}, the location set $\bar{\Phi} = \{y_i\}_{i=0}^\infty$ also forms a homogeneous PPP with spatial density $\lambda$.

We segment the time into equal-length slots with each being the duration to transmit a single packet.
At the beginning of each time slot, every transmitter has the information packets updated according to an independent and identically distributed (i.i.d.) Bernoulli process with parameter $\xi$.{\footnote{While the focus of this paper is on the buffered sources with random packet arrivals, following similar steps in \cite{talak2018optimizing,TriTalMod:19}, the analytical framework can be extended to study the AoI statistics of wireless sensor networks in which packets arrive in periodic patterns.}}
All the incoming packets are stored in a single-server queue with infinite capacity.{\footnote{Although in this paper we adopt infinite-size buffers to ensure all the packets can be ultimately received at the destination, the developed analysis can be extended to explore the AoI performance in networks where the transmitters have finite-size buffers and can discard certain packets \cite{YanAraQue:20Globecom} and further devise optimal packet management schemes for AoI minimization.}}
In this network, we adopt the ALOHA protocol at the transmitters to control the radio channel access. Simply put, during each time slot, the transmitters that have non-empty buffers will send out one packet with probability $p$.
The transmission succeeds if the SINR at the corresponding receiver exceeds a predefined threshold, denoted as $\theta$, upon which the receiver will feedback an ACK and the packet can be removed from the buffer.
Otherwise, the receiver sends a NACK message and the packet is retransmitted in the next available time slot.
We assume the ACK/NACK transmission is instantaneous and error-free, as commonly done in the literature \cite{TalKarMod:18}.
Note that in this system the delivery of packets incurs a delay of one time slot, namely, packets are transmitted at the beginning of time slots and, if the transmission is successful, they are delivered by the end of the same time slot.
Furthermore, we consider two approaches to schedule the packet transmissions:
\begin{itemize}
  \item \textit{FCFS:} The packets are sent out in the order of their arrivals. Once a packet occupies the transmitter, it keeps running until finish.
  \item \textit{LCFS-PR:} A newly arrived packet will stop a currently transmitting but failed one at the end of the time slot, take over the priority and begin its transmission. And only when that finished will the transmitter return to the original task.{\footnote{Note that under the LCFS-PR protocol, packets with the latest timestamps in the queues will be sent out with probability $p$ at the beginning of each time slot, whereas the preempted ones are not discarded but postponed to be transmitted at later time slots (until the packets with the more recent timestamps are successfully received). As pointed out in \cite{ManAbdDhi:20}, such a protocol is optimal in minimizing AoI because it always transmits the latest update available.}}
\end{itemize}
In order to investigate the time-domain evolution, we limit the mobility of transceivers by considering a static network, i.e., the locations of transmitters and receivers remain unchanged in all the time slots \cite{YanQue:19}.

We assume that each transmitter uses unit transmission power $P_{\mathrm{tx}}$.\footnote{ We unify the transmit power to keep the analysis tractable, it shall be noted that the results from this paper can be extended to account for power control via similar approach as in \cite{GhaElsBad:17}. } The channel is subjected to both Rayleigh fading, which varies independently across time slots, and path-loss that follows power law attenuation.
Moreover, the receiver is also subjected to white Gaussian thermal noise with variance $\sigma^2$. By applying Slivnyak's theorem \cite{BacBla:09}, it is sufficient to focus on a \textit{typical} receiver located at the origin, with its tagged transmitter at $X_0$. Thus, when the tagged transmitter sends out a packet during slot $t$, the corresponding SINR received at the typical node takes the following form:
\begin{align} \label{equ:SINR_expression}
\gamma_{0,t} = \frac{P_{\mathrm{tx}} H_{00} r^{-\alpha} }{ \sum_{ j \neq 0 } P_{\mathrm{tx}} H_{j0} \zeta_{j,t} \nu_{j,t} \Vert X_j \Vert^{-\alpha} + \sigma^2 }
\end{align}
where $\alpha$ denotes the path loss exponent, $H_{ji} \sim \exp(1)$ is the channel fading from transmitter $j$ to receiver $i$, $\zeta_{j,t} \in \{ 0, 1 \}$ is an indicator showing whether the buffer of node $j$ is empty ($\zeta_{j,t}=0$) or not ($\zeta_{j,t}=1$), and $\nu_{j,t} \in \{ 0, 1 \}$ represents the channel access decision of node $j$, where it is set to 1 upon assuming transmission approval and 0 otherwise.
\begin{figure}[t!]
  \centering{}

    {\includegraphics[width=0.95\columnwidth]{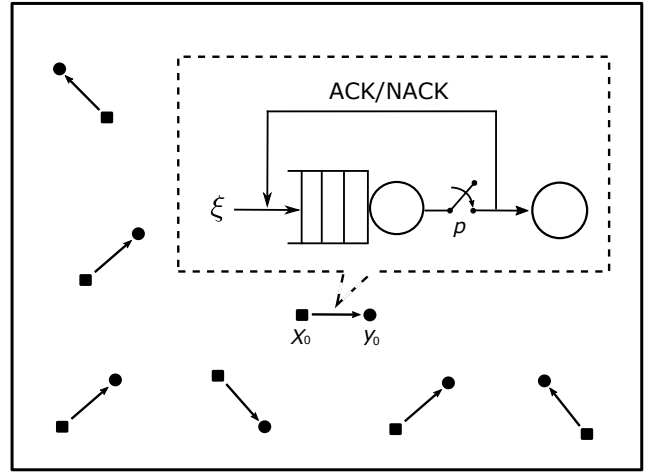}}

  \caption{ An example of the employed network structure, the black squares and dots denote the transmitters and receivers, respectively. Each transmitter has an infinite-size buffer to store all the incoming packets, and switch on with probability $p$ if the buffer is non-empty. }
  \label{fig:NetMod_V1}
\end{figure}

\remark{\textit{The employed system model is particularly relevant to infrastructureless applications like Device-to-Device (D2D) networking, mobile crowd sourcing, and Machine-to-Machine (M2M) communications, that do not require a centralized controller, e.g., bases stations or access points, to coordinate the communications. In fact, such a network is a large-scale analog to the classical model of \textit{Random Networks} \cite{GupKum:00}, in which the distance between any transmitter-receiver pair is fixed to represent the average value. Note that building upon the results from \cite{GhaElsBad:17} and \cite{YanQue:19}, the analysis developed in this paper can be extended to investigate networks with centralized infrastructures and multiple access/broadcast channels where transmitters are located at random distances to their receivers.
}}

\subsection{Age of Information}
Without loss of generality, we denote a randomly selected wireless link $j$ as the generic link and typify the communication link between the transmitter-receiver pair located at $(X_0, y_0)$.
%Then, as illustrated in Figure~\ref{fig:AoIMod_V1}, the AoI over the typical link grows linearly in the absence of successful communication, and, when the transmission is successful, reduces to the time elapsed since the generation of the delivered packet.
%To be concrete, the evolution of $A_0(t)$ can be expressed as follows:
%\begin{align*}
%A_0(t \!+\! 1) =
%\left\{
%       \begin{array}{ll}
%         \!\!  A_0(t)  +   1, \quad \quad ~~   \text{if transmission fails}, \\
%         \!\!  t  -  G_0(t), \quad \quad ~~ \text{otherwise}
%       \end{array}
%\right.
%\end{align*}
%where $G_0(t)$ is the generation time of the packet delivered over the typical link at time $t$.
An example of AoI evolution over the typical link under the LCFS-PR discipline is given in Figure~\ref{fig:AoIMod_V1}. From this figure, we can see that the AoI over the typical link grows linearly in the absence of new packet delivery, and, when the most up-to-date packet is received, reduces to the time elapsed since the generation of the delivered packet.
It is noteworthy that upon successful transmission of a preempted packet, e.g., the one delivered between time slots $t_1$ and $t_2$, the AoI will not be reset because a fresher packet has been delivered at the receiver.
To be more formal, the evolution of $A_0(t)$ can be expressed as follows:
\begin{align*}
A_0(t + 1) =
\left\{
       \begin{array}{ll}
         \!  A_0(t)  +   1, \qquad \qquad ~~~~\,   \text{if transmission fails} \\
         $\qquad$ $\qquad$ $\qquad$ $\qquad$ ~~~~~~  \text{or no transmission}, \\
         \!  \min\{ t  -  G_0(t), A_0(t) \} + 1, \quad \quad ~~ \text{otherwise}
       \end{array}
\right.
\end{align*}
where $G_0(t)$ is the generation time of the packet delivered over the typical link at time $t$.

Due to the randomness from arrival and departure alike, the AoI evolves as a stochastic process. In that respect, we leverage two deterministic quantities, namely the \textit{average} and \textit{peak} AoI, as our metric to evaluate the freshness of information across a wireless network. Specifically, the average AoI at a generic link $j$ is given by
\begin{align}
A^{ \mathrm{ave} }_j = \limsup_{ T \rightarrow \infty } \frac{ 1 }{ T } \sum_{ t=1 }^T A_j(t),
\end{align}
and the peak AoI at link $j$ is defined as
\begin{align} \label{equ:DefPAoI}
A^{ \mathrm{p} }_j = \limsup\limits_{ N \rightarrow \infty } \frac{ \sum_{n=1}^N A_j( T_j(n) ) }{N},
\end{align}
where $T_j(n)$ is the time slot at which the $n$-th packet from link $j$ is successfully delivered. By extending this concept to a large scale, we define the \textit{network} average and peak AoI respectively as follows:
\begin{align}
A^{\mathrm{ave}} &= \limsup_{R \rightarrow \infty} \frac{ \sum_{ X_j \in \tilde{\Phi} \cap B(0,R) } A^{ \mathrm{ave} }_j }{ \sum_{ X_j \in \tilde{\Phi} } \mathbbm{1}\{ X_j \!\in\! B(0,R) \}  }
\nonumber\\
&\stackrel{(a)}{=}   \mathbb{E}^0 \Big[ \limsup_{ T \rightarrow \infty } \frac{1}{T} \sum_{t=1}^T A_0( t )  \Big]
\end{align}
and
\begin{align}
A^{\mathrm{p}} &= \limsup_{R \rightarrow \infty} \frac{ \sum_{ X_j \in \tilde{\Phi} \cap B(0,R) } A^{ \mathrm{p} }_j }{ \sum_{ X_j \in \tilde{\Phi} } \mathbbm{1}{ \{ X_j \!\in\! B(0,R) \} }  }
\nonumber\\
& = \mathbb{E}^0 \Big[ \limsup_{ N \rightarrow \infty } \frac{1}{N} \sum_{n=1}^N A_0( T_0(n) )  \Big],
\end{align}
where $B(0,R)$ denotes a disk centered at the origin with radius $R$, $\mathbbm{1\{ \cdot \}}$ is the indicator function, and $(a)$ follows from the Campbell's theorem \cite{BacBla:09}. The notion $\mathbb{E}^0[\cdot]$ indicates the expectation is taken with respect to the Palm distribution $\mathbb{P}^0$ of the stationary point process -- the condition will be given in Section~III -- where under $\mathbb{P}^0$ almost surely there is a node located at the origin \cite{BacBla:09}.
In what follows, we append the metrics $A^{\mathrm{ave}}$ and $A^{\mathrm{p}}$ with F and L to denote the packets are sent out following FCFS and LCFS-PR disciplines, respectively.
\begin{figure}[t!]
  \centering{}

    {\includegraphics[width=0.95\columnwidth]{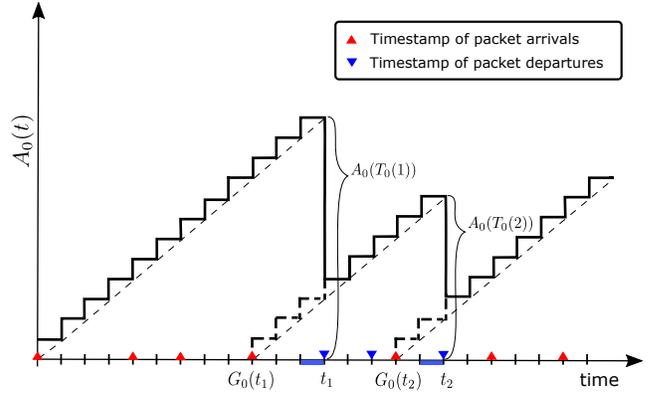}}

  \caption{ AoI evolution example at a typical link under the LCFS-PR discipline. The time instances $G_0(t_i)$ and $t_i$ respectively denote the moment when the $i$-th packet is generated and delivered, and the age is reset to $\min\{ t  -  G_0(t), A_0(t) \} + 1$. Here, $t_i = T_0(i)$ with $T_0(i)$ defined in \eqref{equ:DefPAoI}. }
  \label{fig:AoIMod_V1}
\end{figure}

\subsection{Spatially Interacting Queues}
Because the radio channel is a broadcast medium, transmitters located in proximity can affect each others' queueing states through the interference they cause. As such, the active state $\zeta_{j,t}$ of any given link $j$ is dependent on both space and time, since the former determines the path loss and further the aggregated interference, while the latter affects the queueing process at each node.
%For instance, the service rate in each queue will slow down when the neighboring queues have a larger workload.

To better illustrate this concept, Fig.~\ref{fig:QueItrc_M1} gives a simple example showing the spatiotemporal interactions among the queues of three wireless transmitter-receiver pairs.
From a spatial perspective, we can see that transmitters $X_1$ and $X_2$ are located in geographic proximity and hence their transmissions incur severe crosstalk, which slows down the rate of packet depletion and eventually prolong their queue lengths.
In stark contrast, transmitter $X_3$ locates at a relatively long distance to its geographic neighbors. Such advantage abbreviates transmitter $X_3$ from suffering strong interference
and hence its buffer length is much shorter compared to transmitters $X_1$ and $X_2$.
From a temporal perspective, the traffic load also plays a crucial role in the rate of service and queue length.
Particularly, if all the transmitters have packets to send, the mutual interference will curtail the service rate and prolong the active duration of each transmitter individually. On the other hand, when any of the transmitters silence the packet delivery, the others can benefit from the reduced
cross-talk and speed up their individual queue flushing process, hence also decreases the active period.
Extending this concept to a large-scale network, we find that as the realization of PPP is irregular, there are always some links experiencing poor transmission environment, e.g., those with transmitters located in a crowded area of space, and some others having good communication conditions, e.g., the ones with transmitters that are far away from their neighbors.
Therefore, even the packet update frequencies, or equivalently the arrival rates, are the same for all the transmitters, the queueing status and active state can vary largely from link to link.
To that end, a small variation in the packet update frequency affects not just the inter-arrival rate but also the packet departure process.
This phenomenon is commonly known as the spatially interacting queues and the characterization of its behavior is very challenging.
\begin{figure}[t!]
  \centering{}

    {\includegraphics[width=0.90\columnwidth]{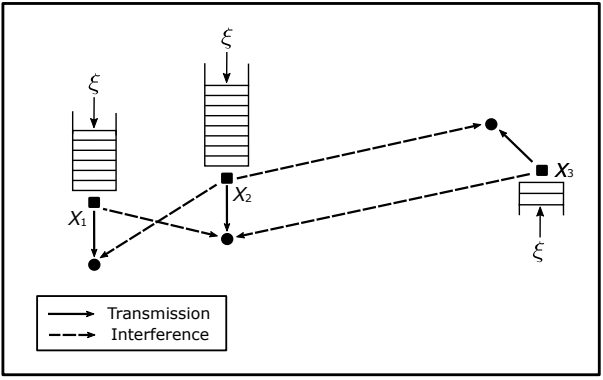}}

  \caption{ Illustration of the spatially queueing interactions. Note that while all the transmitter-receiver pairs are configured with the same distance and update frequencies, their buffer lengths vary due to different levels of interference. }
  \label{fig:QueItrc_M1}
\end{figure}

    \setcounter{equation}{\value{equation}}
    \setcounter{equation}{14}
    \begin{figure*}[t!]
    \begin{align} \label{equ:Meta_Grl}
    F_{\theta}(u) &= \frac{1}{2} -\! \int_{0}^{\infty} \!\!\! \mathrm{Im}\bigg\{ u^{-j\omega} \exp\!\Big(\! - \frac{ j \omega \theta r^\alpha }{ \rho } - 2 \lambda \pi r^2 \sum_{k=1}^{\infty} \binom{j \omega}{ k } (-1)^{k+1} \!\!\int_0^\infty \!\!\! \int_0^{2\pi} \!\! \big[ 1 - \frac{ \xi }{ p \, \mathcal{H}_\theta(v,\psi, p) } \big]^k \frac{d \psi}{2\pi}
    \nonumber\\
    &\qquad \qquad \qquad \qquad \qquad \qquad \qquad \quad \times \int_{0}^{2\pi} \!\Big[\, F_{\theta}\big( \mathcal{H}_\theta(v,\varphi, p) \big) \!+\!\! \int_{ \mathcal{H}_\theta(v,\varphi, p) }^{1}  \!\!\!\!\!\!  \frac{ \mathcal{H}^k_\theta(v,\varphi, p) }{t^k} \, F_{ \theta }(dt) \,
    \Big] \frac{ d \varphi }{ 2\pi }  v dv \Big) \bigg\} \frac{ d \omega }{ \pi \omega }
    \end{align}
    \setcounter{equation}{\value{equation}}{}
    \setcounter{equation}{13}
    \centering \rule[0pt]{18cm}{0.3pt}
    \end{figure*}
    \setcounter{equation}{5}

% ============================================ %
%                 Section: Analysis            %
% ============================================ %
\section{ Analysis }
This section constitutes the main technical part of this paper, in which we derive analytical expressions to characterize the AoI.
Specifically, we analyze the distribution of the conditional transmission success probability, which directly determines the service process.
And based on that we calculate the average and peak AoI in large-scale networks. Unless otherwise stated, the analytical
expressions provided in this section are tight approximations of the transmission success probabilities as well as the AoIs. For better readability, most proofs and mathematical derivations have been relegated to the Appendix.
\subsection{ Preliminaries }
Seen from the temporal perspective, transmissions on any given link can be abstracted as a Geo/G/1 queue where the departure rate is dependent on the communication condition -- which is essentially determined by the geographic structure of the network.
In order to characterize the departure rate, we condition on the realization of point process $\Phi \triangleq \tilde{\Phi} \cup \bar{\Phi}${\footnote{Instead of using the notation $\Phi = \phi$ for the realization, we directly write it as $\Phi$, which is a bit amiss, for the notational simplicity \cite{haenggi2016meta}. Throughout the paper, we sometimes refer to $\Phi$ as a point process, which in fact means the particular realization of the point process $\Phi$.}} and define the conditional transmission success probability at the typical link in a generic time slot $t$ as follows \cite{YanQue:19}
\begin{align}\label{equ:CndTX_Prob}
\mu^\Phi_{0,t} = \mathbb{P}\big(\gamma_{0,t} > \theta | \Phi\big).
\end{align}
Note that $\mu^\Phi_{0,t}$ is a random variable and its distribution is equivalent to the distribution of service rate.
Another observation of \eqref{equ:CndTX_Prob} is that the quantity $\mu^\Phi_{0,t}$ is indexed by the time slot $t$, which implies that the collection of service rates forms a stochastic process over time.
And that may introduce temporal correlations into the queueing process which complicate the subsequent analysis \cite{SanBac:17,SanBacFos:19}.
%According to \eqref{equ:SINR_expression}, the SINR received at each link is dependent on the particular time slot as well as its relative location in the network, which can introduce memory in the queueing process via the spatiotemporal correlations and highly complicate the analysis.
It thus necessitates the introduction of the following approximation.
\begin{assumption}
\textit{
	In this network, each queue observes the time-averages of the activity indicators of other queues but evolves independently of their current state.
}
\end{assumption}

The above assumption makes the dynamics of packet transmissions at each node conditionally independent, given the positions of all transmitters and receivers in the network. Note that this is a mean-field approximation in only the temporal domain, which differs from the mean-field approximation in both space and time, as made in \cite{GhaElsBad:17,YanWanQue:18,YanQue:19,ChiElSCon:17,ChiElSCon:19}. In consequence, we can regard the transmissions of packets over the typical link as i.i.d. over time with a success probability $\mu^\Phi_0 = \lim_{t \rightarrow \infty} \mu^\Phi_0$.
As such, the packet dynamics at the typical transmitter can be abstracted as a Geo/Geo/1 queue, and by leveraging tools from queueing theory, we arrive at a conditional form of the AoI under FCFS.
\begin{lemma} \label{lma:Cnd_AoI_FCFS}
\textit{
	Given $\xi < p \mu^\Phi_{0,\mathrm{F}}$, when conditioned on the point process $\Phi$, the average and peak AoI under FCFS discipline are given respectively as follows:
	\begin{align} \label{equ:Cnd_AoI_FCFS}
	\mathbb{E}^0\!\big[A_{\mathrm{F}}^{\mathrm{ave}} \vert \Phi \big] \!&= \frac{1}{\xi} \!+\! \frac{ 1 - \xi }{ p \mu^\Phi_{0,\mathrm{F}} - \xi } + \frac{ \xi }{ p \mu^\Phi_{0,\mathrm{F}} } - \frac{\xi}{ ( p \mu^\Phi_{0,\mathrm{F}} )^2} - 1, \\ \label{equ:Cnd_PAoI_FCFS}
	\mathbb{E}^0\!\big[A_{\mathrm{F}}^{\mathrm{p}} \vert \Phi \big] \!&= \frac{1}{\xi} \!+\! \frac{ 1 - \xi }{ p \mu^\Phi_{0,\mathrm{F}} - \xi }
	\end{align}
    where $\mu^\Phi_{0,\mathrm{F}}$ denotes the transmission success probability under the considered scenario.
}
\end{lemma}
\begin{IEEEproof}
When conditioned on the point process $\Phi$, the transmission process at a typical link can be regarded as a Geo/Geo/1 queue with the rate of arrival and departure being $\xi$ and $p \mu^\Phi_{0,\mathrm{F}}$, respectively, and hence \eqref{equ:Cnd_AoI_FCFS} and \eqref{equ:Cnd_PAoI_FCFS} follow from leveraging results in \cite{talak2018optimizing}.
\end{IEEEproof}

Similarly, when the packet transmissions are regulated under the LCFS-PR discipline, expressions for the conditional average and peak AoI can be attained accordingly.
\begin{lemma} \label{lma:Cnd_AoI_LCFS}
\textit{
	Given $\xi < p \mu^\Phi_{0,\mathrm{L}}$, when conditioned on the point process $\Phi$, the average and peak AoI under LCFS-PR discipline are given respectively as follows:
	\begin{align} \label{equ:Cnd_AoI_LCFS}
	\mathbb{E}^0\!\big[ A_{ \mathrm{L} }^{\mathrm{ave}} \vert \Phi \big] \!&= \frac{1}{\xi} \!+\! \frac{1}{ p \mu^\Phi_{0,\mathrm{L}} } - 1, \\ \label{equ:Cnd_PAoI_LCFS}
	\mathbb{E}^0\!\big[ A_{ \mathrm{L} }^{\mathrm{p}} \vert \Phi \big] \!&= \frac{ 1 }{\xi} \!+\! \frac{1}{ p \mu^\Phi_{0,\mathrm{L}} } + \frac{1}{1-(1 \!-\! \xi)(1 \!-\! p \mu^\Phi_{0,\mathrm{L}} )} - 2
	\end{align}
    where $\mu^\Phi_{0,\mathrm{L}}$ denotes the transmission success probability under the considered scenario.
}
\end{lemma}
\begin{IEEEproof}
Please see Appendix~\ref{apx:Cnd_AoI_LCFS}.
\end{IEEEproof}

From Lemma~\ref{lma:Cnd_AoI_FCFS} and Lemma~\ref{lma:Cnd_AoI_LCFS}, we can see that the distribution of service rates $\mu^\Phi_{0,\mathrm{F}}$ and $\mu^\Phi_{0,\mathrm{L}}$ plays a role of paramount importance in characterizing the stochastic behavior of AoI.
In that respect, we present a step-by-step derivation of these service rate distributions in the following.

\subsection{ Distribution of Service Rate }
According to \eqref{equ:SINR_expression}, the SINR received at a typical UE comprises a series of random quantities, including the fading, active states and locations of transmitters, which daunts the task of analysis. To launch an initial step, we first average out the randomness from fading and derive a conditional form of the transmission success probability:
\begin{lemma} \label{lma:Cnd_TXS_Prob}
\textit{
    Conditioned on the spatial realization $\Phi$, the transmission success probability at the typical link is given by
    \begin{align} \label{equ:Cnd_TxSuc_Prob}
    \mu^\Phi_{0,z} = e^{ - \frac{\theta r^\alpha}{\rho} } \prod_{j \neq 0} \Big( 1 - \frac{ p \, a^\Phi_{j,z} }{ 1 + D_{j0} } \Big), ~~ z \in \{ \mathrm{F}, \mathrm{L} \}
    \end{align}
    where $\rho = P_{\mathrm{tx}}/\sigma^2$, $D_{ij}=\Vert X_i - y_j \Vert^\alpha / \theta r^\alpha$, and $a_{j,z}^\Phi = \lim_{ t \rightarrow \infty } \mathbb{P}(\zeta_{j,t} = 1 | \Phi, z )$ is the active probability of transmitter $j$ at the steady state.
}
\end{lemma}
\begin{IEEEproof}
By conditioning on the spatial realization $\Phi$ of all the transceiver locations, the transmission success probability can be derived as follows:
\begin{align}
&\mathbb{P} \left( \gamma_0 \!>\! \theta | \Phi  \right)
\!=\! \mathbb{P} \bigg(  { H_{00} }  >  \sum_{ j \neq 0 } \frac{ H_{j0} \nu_j \zeta_{j} \theta r^\alpha }{ \Vert X_j \!-\! y_0 \Vert^\alpha } \!+\! \frac{  \theta r^\alpha }{\rho} \,  \Big| \, \Phi, z  \bigg)
\nonumber\\
&= \mathbb{E}\bigg[ e^{-\frac{ \theta r^\alpha }{\rho}} \prod_{ j \neq 0 } \exp\! \Big(\! - \theta r^\alpha \frac{ \nu_j \zeta_{j} H_{j0} }{ \Vert X_j \!-\! y_0 \Vert^\alpha } \Big)  \Big| \Phi, z \bigg]
\nonumber\\
&\stackrel{(a)}{=}   e^{-\frac{ \theta r^\alpha }{\rho}}  \prod_{j \neq 0} \Big( 1 - p \, a_{j,z}^\Phi  + \frac{ p \, a_{j,z}^\Phi }{ 1 + 1/{D}_{j0} } \Big),
\end{align}
where ($a$) follows by Assumption~1, with which the active state at every transmitter can be treated as independent from each other, and further noticing that $H_{j0} \sim \exp(1)$.
The resultant expression per \eqref{equ:Cnd_TxSuc_Prob} can then be obtained by further simplifying the product factors.
\end{IEEEproof}

The result from Lemma~\ref{lma:Cnd_TXS_Prob} explicitly reveals that the randomness associated with the transmission success probability is mainly assorted to $i$) the random location of the interfering transmitters, and $ii$) their corresponding active states.
As such, by leveraging the queueing theory, a conditional expression for the active state at each transmitter can be obtained as follows.
\begin{lemma} \label{lam:actv_prob}
\textit{
    Given the service rate $\mu^\Phi_j$, the queue-nonempty probability at a generic transmitter $j$ is given as
    \begin{align}\label{equ:ActProb_Gnrl}
    a^\Phi_j = \left \{
    \begin{tabular}{cc}
    \!\!\!\!\!\text{1}, & \text{if}~ $ p \, \mu^\Phi_j \leq \xi$,   \\
    % ~\\
    \!\!\!\!  $\frac{ \xi }{ p \mu^\Phi_j }$, & \text{if}~ $ p \, \mu^\Phi_j >  \xi$.
    \end{tabular}
    \right.
    \end{align}
}
\end{lemma}
\begin{IEEEproof}
For a Geo/G/1 queue, the probability of being active in the steady state follows from the Little's law \cite{Har:13}.
\end{IEEEproof}

Notably, the result in Lemma~\ref{lam:actv_prob} applies to any transmitter of the depicted network regardless of the specific scheduling method it employed. The immediate aftermath of this observation is the following:
\begin{align}
\mu^\Phi_{0,\mathrm{F}} \stackrel{ d }{=} \mu^\Phi_{0,\mathrm{L}},
\end{align}
namely the conditional transmission success probabilities under LCFS-PR and FCFS policies are equivalent in distribution.
On this account, we drop the index of scheduling policy and simply write $\mu^\Phi_{0}$ in lieu of $\mu^\Phi_{0,z}$.
By substituting \eqref{equ:ActProb_Gnrl} into \eqref{equ:Cnd_TxSuc_Prob}, we also note that all the service rates are intertwined with each other. In fact, an increase of any one of the $\mu^\Phi_j$ can also boost up the value of others and vice versa.
To capture these interactions, we decompose the derivation of the distribution into two major steps: $i$) conditioned on the distance between the typical receiver and a generic receiver, calculate the probability of each node being in the active state, and then $ii$) compute the distribution of the conditional transmission success probability.
That brings us to the first technical result.
\begin{theorem} \label{thm:Meta_SINR}
  \textit{
  The cumulative distribution function (CDF) of the conditional transmission success probability is given by the fixed-point equation \eqref{equ:Meta_Grl} at the top of this page, in which $\mathrm{Im}\{ \cdot \}$ denotes the imaginary part of a complex quantity  and the auxiliary function $\mathcal{H}_\theta(x,y,z)$ is given by
    \setcounter{equation}{\value{equation}}
    \setcounter{equation}{15}
  \begin{align} \label{equ:H_theta}
  \mathcal{H}_\theta(x,y,z) = \frac{\xi}{z} + \frac{\xi}{ 1 - z + (1+x^2-2x \cos y)^{ \frac{\alpha}{2} }/\theta }.
  \end{align}
  Furthermore, \eqref{equ:Meta_Grl} can be solved via recursive calculations as follows:
  \begin{align}
  F_\theta(u) = \lim_{ n \rightarrow \infty } F_{ \theta, n }(u)
  \end{align}
  where $F_{ \theta, n }(u)$ is given by
  \begin{align} \label{equ:F_theta_n}
  F_{ \theta, n }(u) &= \frac{1}{2} - \! \int_0^\infty \!\!\!\!\! \mathrm{Im} \bigg\{ u^{-j\omega} \exp\!\Big(\! - \frac{ j \omega \theta r^\alpha }{ \rho }
  \nonumber\\
  & \quad \quad - \frac{ \lambda r^2 }{ 2 \pi } \sum_{k=1}^{\infty} \binom{j \omega}{ k } (-1)^{k+1} \tilde{\eta}^{(k)}_{n-1} \Big) \bigg\} \frac{ d \omega }{ \pi \omega },
  \end{align}
  in which $\tilde{\eta}^{(k)}_{n-1}$ takes the following form:
  \begin{align}
  \tilde{\eta}^{(k)}_{n-1} & =\!\! \int_0^\infty \!\!\!\! \int_0^{ 2 \pi } \!\!\!\!\! \big[ 1 \!-\! \frac{ \xi }{ p \mathcal{H}_\theta( v, \psi, p ) } \big]^k \!\! d \psi \! \! \int_{0}^{2\pi}  \!\!\!\!\!\! \big[\, F_{ \theta, n-1 }\big( \mathcal{H}_\theta( v, \varphi, p ) \big)
\nonumber\\
& \quad + \!\! \int_{ \mathcal{H}_\theta( v, \varphi, p ) }^{ 1 } \!\!\!\!  \frac{ \mathcal{H}^k_\theta( v, \varphi, p ) }{t^k} F_{ \theta, n-1 }( dt ) \big] d \varphi v dv.
  \end{align}
  In particular, when $n=1$, we have $\tilde{\eta}_0^{(k)}$ given as
  \begin{align}
  \tilde{\eta}_0^{(k)} = \binom{ \delta - 1 }{ k - 1 } \frac{ 2 \pi^2 \delta \theta^\delta \xi^k p^k }{ \sin( \pi \delta ) }.
  \end{align}
    \setcounter{equation}{\value{equation}}{}
    \setcounter{equation}{20}
    \setcounter{equation}{20}
  }
\end{theorem}
\begin{IEEEproof}
Please see Appendix~\ref{apx:Meta_SINR}.
\end{IEEEproof}

A marked distinction between the CDF given in \eqref{equ:Meta_Grl} and those obtained via the mean-field approximation \cite{GhaElsBad:17,YanWanQue:18,YanQue:19,ChiElSCon:17,ChiElSCon:19} is that the former represents the distribution of conditional transmission success probability, a.k.a. SINR meta distribution \cite{haenggi2016meta}, of a \textit{non-homogeneous} PPP.
As already illustrated in Section~II-C, the spatial queueing interactions amongst the transmitters lead to a location-dependent active pattern at each node. Therefore,
seen from the perspective of a typical transmitter, the interferers, i.e., the nodes that are activated at the same time, are by nature non-homogeneous in their spatial distributions and it is important to account for this phenomenon in the analysis.

However, directly computing $F_\theta(u)$ requires all moments of $\mu^\Phi_0$, which is time-consuming. One way to get around this is via approximation.

\begin{corollary}\label{Cor:Approximation}
\textit{The probability density function (pdf) of $F_{\theta}(u)$ in Theorem~\ref{thm:Meta_SINR} can be tightly approximated via the following
\begin{align} \label{equ:fX}
f_X(u) &= \lim_{ n \rightarrow \infty } f_{X_n}(u)
\nonumber\\
&= \lim_{ n \rightarrow \infty } \frac{u^{\frac{\kappa_n (\beta_n + 1) - 1}{1 - \kappa_n}} (1-u)^{\beta_n - 1} }{B(\kappa_n \beta_n/(1-\kappa_n), \beta_n )}
\end{align}
where $B(a, b)$ denotes the Beta function \cite{AndAsk:00}, $\kappa_n$ and $\beta_n$ are respectively given as
\begin{align} \label{equ:mu_n}
\kappa_n &= c_n^{(1)}, \\ \label{equ:beta_n}
\beta_n &= \frac{ ( 1 - \kappa_n ) \big[\kappa_n - c_n^{(2)} \big] }{c_n^{(1)} - \kappa_n^2 }
\end{align}
where $c_n^{(m)}$ can be written as
\begin{align} \label{equ:Momnt_Beta}
c_n^{(m)} \!=\! \exp\!\bigg(\!\! -\! \frac{ m \theta r^\alpha }{ \rho }  \! - \! \lambda r^2 \! \sum_{k=1}^{m} \! \binom{ m }{ k } (-1)^{ k+1 } \, \hat{\eta}^{(k)}_n \!\bigg),
\end{align}
and $\hat{\eta}^{(k)}_n$ is given by
\begin{align}
\hat{\eta}^{(k)}_{n-1} & =\!\! \int_0^\infty \!\!\! \int_0^{ 2 \pi } \!\!\!\! \big[\, 1 - \frac{ \xi }{ p \mathcal{H}_\theta( v, \psi, p ) } \, \big]^k d \psi \int_{0}^{2\pi}  \!\!\!\! \big[\! \int_{0}^{ \mathcal{H}_\theta( v, \varphi, p ) } \!\!\!\!\!\!\!\!\!\!\!\!\!\!\!\!\!\!   f_{ X_{n - 1} }(t)dt
\nonumber\\
& \qquad + \int_{ \mathcal{H}_\theta( v, \varphi, p ) }^{ 1 } \!\!\!\!\!   \frac{ \mathcal{H}^k_\theta( v, \varphi, p ) }{t^k} f_{ X_{n - 1} }(t)  dt \big] \frac{d \varphi}{ 2 \pi } v dv.
\end{align}
Particularly, when $n=1$, we have $\hat{\eta}_{0}^{(k)}$ given by the following
\begin{align}
\hat{\eta}_{0}^{(k)} = \binom{ \, \delta - 1 \,}{\, k - 1 \,} \frac{ 2 \pi^2 \theta^{ \delta } \xi^k p^k }{ \alpha \sin(  \pi \delta ) }.
\end{align}
}
\end{corollary}
\begin{IEEEproof}
It can be observed from \eqref{equ:Meta_Grl} that the approximated function $F_{\theta,n}(u)$ in each iteration step is supported on $[0,1]$. We are thus motivated to approximate the distribution via a Beta distribution.
First, by assigning $s$ as integers as per \eqref{equ:MG_MY0}, we can derive the moments in \eqref{equ:Momnt_Beta}. Next, by respectively matching the mean and variance to a Beta distribution $B(a_n, b_n)$, it yields
\begin{align}
& \frac{ a_n }{ a_n + b_n } = c_n^{(1)}, \\
& \frac{ a_n b_n }{ ( a_n + b_n )^2 ( a_n + b_n + 1 ) } = c_n^{(2)} - \big[ c_n^{(1)} \big]^2
\end{align}
and the result follows from solving the above system equations.
\end{IEEEproof}

Note that compared to \eqref{equ:Meta_Grl}, calculation of \eqref{equ:fX} involves only the first two moments of $\mu^\Phi_0$ and hence can be efficiently executed.
When conducting the computation, we generally set a sufficiently small threshold  $\epsilon$ and stop the iteration when $\vert \hat{\eta}_n^{(k)} - \hat{\eta}_{n-1}^{(k)} \vert < \epsilon, \forall k$ where $\hat{\eta}_n^{(k)}$ is given in (25). It has been shown that such a iteration can converge in very few, e.g., less than 10, steps \cite{ZhaYanShe:20}.
The accuracy of the above derivations will be verified in Fig.~\ref{fig:SimVerf} in Section~IV.

Armed with these results, we are now in a position to give a complete characterization of the AoI.

\subsection{ Stable Region and AoI }
Before delving into the calculation of AoI, we would like to pause and present the condition under which the queueing network is stable, i.e., a typical transmitter does not explode its buffer \cite{BacRybShl:18}. This task can be accomplished by controlling the update frequency at each communication link:
\begin{theorem} \label{thm:stability}
\textit{The sufficient and necessary conditions for the queueing network to remain stable can be tightly approximated as follows:
\begin{align} \label{equ:Stb_Regn}
\xi \leq \xi_{ \mathrm{c} } = \sup \{ \xi | \xi \leq p \cdot p_{ \mathrm s } \}
\end{align}
where $\xi_{ \mathrm{c} }$ is the critical update frequency, and $p_{\mathrm s}$ is the probability of success transmission at the typical node, given as follows
\begin{align} \label{equ:ExcForm_ps}
& p_{\mathrm{s}} = \exp\!\Big(\! -\! \frac{ \theta r^\alpha }{ \rho } \!-\! \lambda r^2\!\! \int_0^\infty \!\!\!\! \int_0^{2\pi} \!\!\!\!\! \frac{ \mathcal{Z}_\theta(v,p_{\mathrm{s}}, p, \xi) v d \varphi dv }{ 1 \!+\! ( 1 \!+\! v^2 \!-\! 2 v \cos \varphi )^{ \frac{\alpha}{2} } \!/ \theta } \Big) \\
& \approx \exp\!\Big(\! - \! \frac{ \theta r^\alpha }{ \rho } - \lambda r^2 \theta^{\frac{2}{\alpha}} \!\!\! \int_0^\infty  \frac{ \min \big\{ \frac{\xi}{p_{\mathrm{s}} }( 1 \!+\! u^{-\frac{\alpha}{2}} ), p \big\} }{ 1 + u^{ \frac{\alpha}{2} } } du \Big)
\end{align}
where $\mathcal{Z}_{\theta}(v,p_{\mathrm{s}}, p, \xi)$ is given by
\begin{align}
\mathcal{Z}_{\theta}(v,p_{\mathrm{s}}, p, \xi) \!=\!\! \int_0^{2\pi} \!\!\!\!\!\!\! \min\!\Big\{ \frac{\xi}{p_{\mathrm{s}}}\big[ 1 \!+\! \frac{ \theta }{ (1 \!-\! 2 v \cos \psi \!+\! v^2 )^{\frac{\alpha}{2}} }  \big], p \Big\} \frac{ d \psi }{ 2 \pi }.
\end{align}
}
\end{theorem}
\begin{IEEEproof}
Please see Appendix~\ref{apx:stability}.
\end{IEEEproof}
In the sequel, we restrict the value of network parameters to be set within the stable region such that the resultant statistics are well defined.

At this stage, we are ready to finally derive expressions for the AoI.
\begin{theorem} \label{thm:AoI_FCFS}
\textit{
	Under the FCFS discipline, the average and peak AoI are given as follows:
	\begin{align} \label{equ:EctForm_AoI_FCFS}
	A_{\mathrm{F}}^{\mathrm{ave}} &= \frac{1}{\xi} + \!\! \int_{\xi/p}^{1} \! \Big(\, \frac{ 1 - \xi }{ p t - \xi }  +  \frac{ \xi }{ p t } - \frac{\xi}{ p^2 t^2} \,\Big) F_{\theta}(dt) -1
    \nonumber \\
    &\approx \frac{1}{\xi} + \!\! \int_{\xi/p}^{1} \! \Big(\, \frac{ 1 - \xi }{ p t - \xi } + \frac{ \xi }{ p t } - \frac{\xi}{ p^2 t^2}  \,\Big) f_X(t) dt - 1, \\ \label{equ:EctForm_PAoI_FCFS}
	A_{\mathrm{F}}^{\mathrm{p}} &= \frac{1}{\xi} + \! \int_{\xi/p}^{1} \! \Big(\, \frac{ 1 - \xi }{ p t - \xi } \,\Big) F_{\theta}(dt)
    \nonumber \\
    &\approx \frac{1}{\xi} + \! \int_{\xi/p}^{1} \! \Big(\, \frac{ 1 - \xi }{ p t - \xi } \,\Big) f_X(t) dt
	\end{align}
    where $F_\theta(\cdot)$ and $f_{X}(\cdot)$ are given in \eqref{equ:Meta_Grl} and \eqref{equ:fX}, respectively.
}
\end{theorem}
\begin{IEEEproof}
The above expressions are attained by deconditioning the random variable $\mu^\Phi_0$ in \eqref{equ:Cnd_AoI_FCFS} and \eqref{equ:Cnd_PAoI_FCFS} with respect to its distribution functions given per \eqref{equ:Meta_Grl} and \eqref{equ:fX}, respectively.
\end{IEEEproof}

\begin{theorem} \label{thm:AoI_LCFS}
\textit{
	Under the LCFS-PR discipline, the average and peak AoI are given as follows:
	\begin{align} \label{equ:EctForm_AoI_LCFS}
	A_{\mathrm{L}}^{\mathrm{ave}} &= \frac{1}{\xi} -1 + \!\! \int_{\xi/p}^{1} \!\! \frac{ F_{\theta}(dt) }{ p t }
    \nonumber \\
    &\approx \frac{1}{\xi} -1 +  \!\! \int_{\xi/p}^{1} \!\! \frac{  f_X(t) dt }{ p t }, \\ \label{equ:EctForm_PAoI_LCFS}
	A_{\mathrm{F}}^{\mathrm{p}} &= \frac{1}{\xi} - 2 + \!\! \int_{\xi/p}^{1} \!\! \big[\, \frac{1}{pt} + \frac{ 1 }{ 1 - (1-\xi)( 1 - \xi t) } \,\big] F_{\theta}(dt)
    \nonumber \\
    &\approx \frac{1}{\xi} - 2 + \!\! \int_{\xi/p}^{1} \!\! \big[\, \frac{ f_X(t) }{pt} + \frac{ f_X(t) }{ 1 - (1-\xi)( 1 - \xi t) } \,\big] dt
	\end{align}
    where $F_\theta(\cdot)$ and $f_{X}(\cdot)$ are given by \eqref{equ:Meta_Grl} and \eqref{equ:fX}, respectively.
}
\end{theorem}
\begin{IEEEproof}
This result can be obtained via similar approaches in Theorem~\ref{thm:AoI_FCFS} and hence is omitted here.
\end{IEEEproof}

To this end, \eqref{equ:Meta_Grl} quantifies not only the key features of a wireless network, including the deployment density, interference, and traffic dynamics, but also the impact of spatially interacting queues, via a fixed-point functional equation.
%Several remarks regarding Theorem~\ref{thm:AoI_FCFS} and Theorem~\ref{thm:AoI_LCFS} are in order.
%\remark{\textit{By comparing \eqref{equ:EctForm_AoI_FCFS} and \eqref{equ:EctForm_AoI_LCFS}, we have the following:
%\begin{align}
%A^{\mathrm{ave}}_{\mathrm{F}} - A^{\mathrm{ave}}_{\mathrm{L}} = 1 +\! \int_{\xi/p}^{1} \!\frac{ (1-pt) \xi^2 }{(pt-\xi)p^2 t^2 } F_\theta(dt) > 1,
%\end{align}
%which implies the average AoI attained under the LCFS-PR discipline is always smaller than that under the FCFS discipline, and the difference is lower bounded by one.
%}}
%
%\remark{\textit{By comparing \eqref{equ:EctForm_PAoI_FCFS} and \eqref{equ:EctForm_PAoI_LCFS}, we note that there exist $p>0$ such that when $\xi \ll 1$, the following holds:
%\begin{align}
%A^{\mathrm{p}}_{\mathrm{F}} - A^{\mathrm{p}}_{\mathrm{L}} = {o}(\xi),
%\end{align}
%i.e., the difference between the peak AoI attained under the LCFS-PR and FCFS discipline can be arbitrarily small.
%}}

To further release the burden on computing \eqref{equ:EctForm_AoI_FCFS} to \eqref{equ:EctForm_PAoI_LCFS}, the following corollaries provide low-complexity approximations.
\begin{corollary} \label{cor:aprx_AoI_FCFS}
\textit{
	Under the FCFS discipline, the average and peak AoI can be respectively approximated as follows:
	\begin{align} \label{equ:aprx_AoI_FCFS}
	A_{\mathrm{F}}^{\mathrm{ave}} &\approx \frac{1}{\xi} + \frac{ 1 - \xi }{ p \cdot p_{\mathrm s} - \xi } + \frac{ \xi }{ p \cdot p_{\mathrm s} } - \frac{\xi}{ p^2 \cdot p_{\mathrm s}^2} - 1, \\ \label{equ:aprx_PAoI_FCFS}
	A_{\mathrm{F}}^{\mathrm{p}} &\approx \frac{1}{\xi} + \frac{ 1 - \xi }{ p \cdot p_{\mathrm s} - \xi }
	\end{align}
    where $p_{\mathrm s}$ is given by \eqref{equ:ExcForm_ps}.
}
\end{corollary}
\begin{IEEEproof}
By deconditioning $\mu^\Phi_{0,\mathrm{F}}$ in \eqref{equ:Cnd_AoI_FCFS}, we have
\begin{align}
& A_{\mathrm{F}}^{\mathrm{ave}} \!= \frac{1}{\xi} \!+\! \mathbb{E}^0\Big[\, \frac{ 1 - \xi }{ p \mu^\Phi_{0,\mathrm{F}} - \xi } + \frac{ \xi }{ p \mu^\Phi_{0,\mathrm{F}} } - \frac{\xi}{ ( p \mu^\Phi_{0,\mathrm{F}} )^2} \, \Big]
\nonumber\\
&\approx \frac{1}{\xi} \!+\! \frac{ 1 - \xi }{ p \mathbb{E}^0\big[ \mu^\Phi_{0,\mathrm{F}} \big]- \xi } + \frac{ \xi }{ p \mathbb{E}^0\big[ \mu^\Phi_{0,\mathrm{F}} \big] } - \frac{\xi}{ ( p \mathbb{E}^0\big[ \mu^\Phi_{0,\mathrm{F}} \big] )^2},
\end{align}
and then \eqref{equ:aprx_AoI_FCFS} follows. Analogously, we can obtain \eqref{equ:aprx_PAoI_FCFS} via the same argument.
\end{IEEEproof}

\begin{corollary}
\textit{
	Under the LCFS-PR discipline, the average and peak AoI can be respectively approximated as follows:
	\begin{align} \label{equ:aprx_AoI_LCFS}
	A_{ \mathrm{L} }^{\mathrm{ave}} &\approx \frac{1}{\xi} + \frac{1}{ p \cdot p_{ \mathrm{s} } } - 1, \\ \label{equ:aprx_PAoI_LCFS}
	A_{ \mathrm{L} }^{\mathrm{p}} &\approx \frac{ 1 }{\xi} + \frac{1}{ p \cdot p_{\mathrm s} } + \frac{1}{1-(1-\xi)(1 - p \cdot p_{\mathrm s} )} - 2
	\end{align}
    where $p_{\mathrm s}$ is given by \eqref{equ:ExcForm_ps}.
}
\end{corollary}
\begin{IEEEproof}
This result can be obtained via similar approaches in Corollary~\ref{cor:aprx_AoI_FCFS} and hence is omitted here.
\end{IEEEproof}

Note that according to the Jensen's inequality, the results in \eqref{equ:aprx_PAoI_FCFS}, \eqref{equ:aprx_AoI_LCFS}, and \eqref{equ:aprx_PAoI_LCFS}, as well as the second and third terms of \eqref{equ:aprx_AoI_FCFS}, are in fact lower bounds of the original expressions. And such a lower bound has been adopted in many stochastic geometry related network analysis as a tight approximation \cite{WanZhoRee:13,ChaRyuQue:15}.

\subsection{Special Case Study }

In light of the developed analysis, we explore in this section the AoI statistic under several special cases to gain further insights.
Specifically, we investigate three scenarios, in which the network is operating at an interference level that is above, below, and equal to, the original system, respectively.

\subsubsection{Dominant System Scenario} In a dominant system, every transmitter except the typical one is backlogged, i.e., only the typical link has packet dynamics over it while the interferers are alway having packets to be sent out. Consequently, the typical transmitter experiences a higher level of interference than that in the original system. We denote the conditional transmission success probability of the typical link in such a system as $\hat{\mu}^\Phi_0$, and the average throughput of the typical link can then be calculated as follows:
\begin{align}\label{equ:ThrPut_DmnSym}
p \cdot \mathbb{E}\left[ \hat{\mu}^\Phi_0 \right] &\stackrel{(a)}{=} p \cdot \mathbb{E}\bigg[ e^{- \frac{\theta r^\alpha}{\rho} } \prod_{ j \neq 0 } \Big( 1 - \frac{ p }{ 1 + D_{j0} } \Big) \bigg]
\nonumber\\
&\stackrel{(b)}{=} p \cdot \exp \left( - \frac{ \theta r^\alpha }{ \rho } - \lambda \pi r^2 \theta^\delta \!\! \int_0^\infty \!\!\! \frac{ p \, dv }{ 1 + v^{\alpha / 2} } \right)
\end{align}
where ($a$) follows from assigning $a^\Phi_j = 1, \forall j \neq 0$ in \eqref{equ:Cnd_TxSuc_Prob} and ($b$) by leveraging the PGFL of PPP to carry out the calculation. Then, by taking a derivative of the throughput in \eqref{equ:ThrPut_DmnSym} with respect to $p$, and setting it to be zero, i.e., $\frac{\partial (p \cdot \mathbb{E} [ \hat{\mu}^\Phi_0 ] ) }{\partial p} = 0$, we can solve for the optimal channel access probability as the following:
\begin{align}\label{equ:OptpStr}
p^* = \min\Big\{ \frac{1}{ \lambda \pi r^2 \theta^\delta \int_0^\infty \frac{ dv }{ 1 + v^{ \alpha/2 } }  }, 1 \Big\}.
\end{align}
Following \eqref{equ:OptpStr}, it is clear that under the dominant system, there may exist a channel access probability $p^* \in (0, 1)$ that optimizes the throughput, which, according to Corollaries 2 and 3, then in turn minimizes the average and peak AoI under FCFS and LCFS-PR.

\subsubsection{Sparsely Deployed Networks} If the network is sparsely deployed, namely, $\lambda \rightarrow 0$, the transmitting nodes recede into the distance and the interference becomes negligible. Therefore, according to \eqref{equ:ExcForm_ps}, we have $p_{\mathrm{s}} \approx e^{-\frac{ \theta r^\alpha }{ \rho } }$. As such, following the Corollaries~2 and 3, we can approximate the average and peak AoI under FCFS and LCFS-PR protocols respectively as follows:
\begin{align}
A^{ \mathrm{ave} }_{\mathrm{F}} &\approx \frac{1}{\xi} + \frac{ 1 - \xi }{ p \cdot e^{- \frac{\theta r^\alpha}{\rho} } - \xi } + \frac{ \xi e^{ \frac{\theta r^\alpha }{ \rho } } }{ p } - \frac{ \xi e^{ \frac{ 2 \theta r^\alpha }{ \rho } } }{ p^2 } - 1, \\
A^{ \mathrm{p} }_{\mathrm{F}} &\approx \frac{1}{\xi} + \frac{ 1 - \xi }{ p e^{-\frac{\theta r^\alpha }{\rho}} - \xi }, \\
A^{ \mathrm{ave} }_{\mathrm{L}} &\approx \frac{1}{\xi} + \frac{e^{\frac{\theta r^\alpha}{\rho} } }{ p } - 1, \\
A^{ \mathrm{p} }_{\mathrm{L}} &\approx \frac{1}{\xi} + \frac{e^{ \frac{\theta r^\alpha }{\rho} } }{p} + \frac{1}{ 1 - (1-\xi) (1 - p e^{-\frac{ \theta r^\alpha }{ \rho }} ) } - 2.
\end{align}
From the above results, we can see that the AoI decreases monotonically with respect to the channel access probability $p$. Moreover, if we set $p = 1$, and keep reducing the decoding threshold, i.e., $\theta \rightarrow 0$, there is $A^{ \mathrm{ave} }_{\mathrm{F}} \rightarrow A^{ \mathrm{ave} }_{\mathrm{L}}$, namely the packet transmission protocols have a mild effect on the performance of AoI when the wireless links have very high throughput.

\subsubsection{Spatially Interacting Queues} In the presence of spatially queueing interactions as the depicted network, we can take a derivative of the average throughput, $p \cdot p_{ \mathrm{s} }$, of the typical link with respect to $p$ and have the following
\begin{align} \label{equ:SptQ_Derv}
\frac{ \partial (p \cdot p_{ \mathrm{s} } ) }{ \partial p } \!&=\! \bigg(\! 1 - p \lambda r^2\!\! \int_0^\infty \!\!\!\! \int_0^{2\pi} \!\!\!\!\! \frac{ \frac{ \partial \mathcal{Z}_\theta(v,p_{\mathrm{s}}, p, \xi) }{ \partial p } v d \varphi dv }{ 1 \!+\! ( 1 \!+\! v^2 \!-\! 2 v \cos \varphi )^{ \frac{\alpha}{2} } \!/ \theta }
 \bigg) \!\! \times \! p_{ \mathrm{s} }\\ \label{equ:SptQ_Derv_LwB}
& \geq \! \left(\! 1 - p \lambda r^2\!\! \int_0^\infty \!\!\!\! \int_0^{2\pi} \!\!\!\!\! \frac{ v d \varphi dv }{ 1 \!+\! ( 1 \!+\! v^2 \!-\! 2 v \cos \varphi )^{ \frac{\alpha}{2} } \!/ \theta } \right) \!\! \times\! p_{ \mathrm{s} }.
\end{align}

From \eqref{equ:SptQ_Derv_LwB}, we note that for sparse network deployment, i.e., $\lambda$ is relatively small, there is $\frac{ \partial (p \cdot p_{ \mathrm{s} } ) }{ \partial p } \geq 0$ and hence the link throughput keeps increasing with channel access probability $p$ which implies the AoI always decreases with $p$. On the other hand, when the network is densely deployed, namely $\lambda$ is relatively large, \eqref{equ:SptQ_Derv} may be negative for a large $p$ value. In that context, there may exist an optimal channel access probability $p \in (0, 1)$ that maximizes the link throughput and so as the AoI. Additionally, by comparing \eqref{equ:ThrPut_DmnSym} and \eqref{equ:SptQ_Derv}, one shall note that the optimal $p$ in networks with spatially interacting queues are always greater than that of a dominant system.

% ============================================ %
%         Section: Sim & Num Analysis          %
% ============================================ %
\begin{figure}[t!]
  \centering{}

    {\includegraphics[width=0.95\columnwidth]{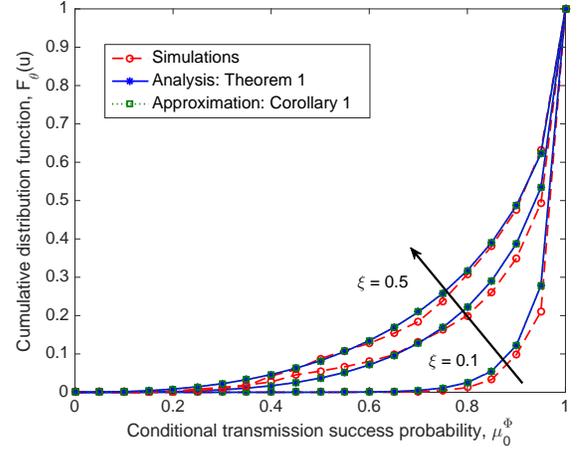}}

  \caption{ Simulation versus analysis: CDF of the conditional transmission success probability, in which we set $r=25$ m, and vary the packet update frequencies as $\xi = 0.1, 0.3, 0.5$. }
  \label{fig:SimVerf}
\end{figure}
\section{Simulation and Numerical Results}
In this section, we verify the accuracy of our analysis through simulations and then evaluate the performance of AoI under different sets of network parameters.
Particularly, during each simulation run, we realize the locations of the transmitters and receivers over a 5~$\text{km}^2$ area via independent PPPs. The packets of each node are updated according to independent Bernoulli processes.
We average over 10,000 realizations and collect the statistic from each communication link to finally calculate the corresponding metrics.
Unless differently specified, we use the following parameters: $\alpha = 3.8$, $\xi=0.3$, $\theta=0$~dB, $P_{\mathrm{tx}}=17$~dBm, $\sigma^2 = -90$~dBm, $p=0.6$, $r=15$ m and $\lambda = 10^{-4}\mathrm{m}^{-2}$.

In Fig.~\ref{fig:SimVerf}, we compare the simulated CDF of the conditional transmission success probability to the analysis developed in Theorem~1 and the approximations in Corollary~1, for various values of packet update frequency $\xi$.
The figure shows a close match between the analytical results and simulations, thus confirms the accuracy of the theorem.
Besides, the differences between the analysis in \eqref{equ:Meta_Grl} and approximation per \eqref{equ:fX} are almost indistinguishable, which verifies the tightness of the approximation.
We also observe that the conditional transmission success probability monotonically decreases with the increase of packet update frequency, because more and more wireless links are activated and that rise up the interference level across the network.
Moreover, the increase in the update frequency defects the probability of transmission success in a non-linear manner, whereas the degradation of transmission success probability is more severe as the updating frequency goes from light ($\xi = 0.10$) to medium ($\xi = 0.30$), and the decreasing trend slows down as the network load further increases to heavy traffic regime ($\xi = 0.50$).
The reason comes from the composite effect of the updating rate.
In the light traffic condition, as the packet arrival rate goes up, the increased traffic load not only wakes up more transmitters but also brings in more accumulated packets at the buffer.
Together with the reduced service rate, the active duration of transmitters is extended, which in turns defect the SINR across the network.
In the heavy traffic regime, as most of the queues are already saturated, the additional active links cannot largely change the interference, and thus the SINR coverage probability descent is leveled off.

\begin{figure}[t!]
  \centering{}

    {\includegraphics[width=0.95\columnwidth]{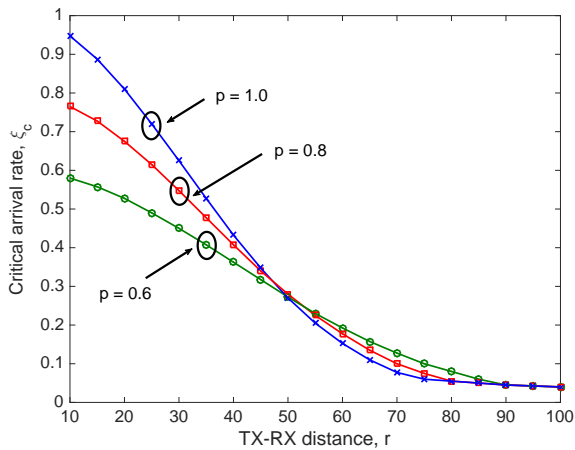}}

  \caption{ Stable region: critical update frequency vs distance between a generic transmitter-receiver pair. }
  \label{fig:StblRegn}
\end{figure}

Fig.~\ref{fig:StblRegn} provides the conditions under which the queueing network can remain stable. In this figure, the critical update frequency is depicted as a function of the distance between a generic transmitter and its intended receiver. We can see that the maximally allowable update frequency declines dramatically as the transceiver distance increases since that reduces the transmission success probability.
Moreover, we can see that after $r=50$ -- the midpoint of the average inter-site distance -- the order of the critical update frequency reverses, because in this regime interference dominates the transmission and suppressing that can benefit the transmitters.

\begin{figure}[t!]
  \centering{}

    {\includegraphics[width=0.95\columnwidth]{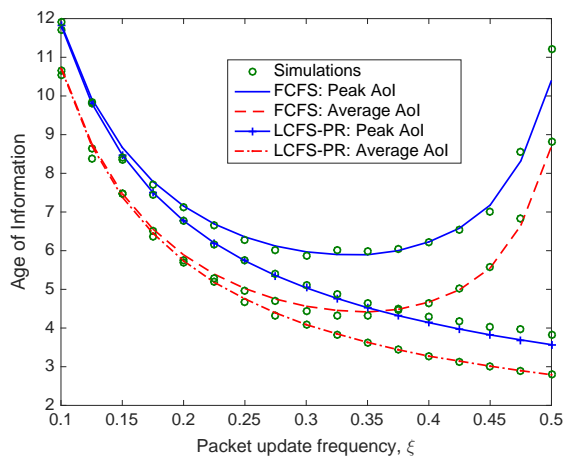}}

  \caption{ Comparing the impact of transmission protocols on the performance of AoI. }
  \label{fig:AoI_vs_xi}
\end{figure}

Fig.~\ref{fig:AoI_vs_xi} shows the simulation and analytical results of both ($a$) the peak and ($b$) the average AoI for a varying value of the packet update frequency. We immediately notice that the simulations and analyses match well with each other, which validates the accuracy of Theorem~3 and Theorem~4. Moreover, we observe that under the FCFS discipline, there exists an optimal update frequency that minimizes both the average and peak AoI, due to the tradeoff between the aggressiveness of updating and the overall interference level that affects the transmission success probability. In stark contrast, we can see that networks operating under LCFS-PR are able to attain smaller AoI than those under FCFS, and both the peak and average AoI decline monotonically with respect to the update frequency.
These observations also coincide with those drawn from the abstract models \cite{CosCodEph:16,Yat:18}, thus confirm the effectiveness of LCFS-PR over FCFS in reducing AoI from of perspective of an SINR model.

%\begin{figure}[t!]
%  \centering{}
%
%    {\includegraphics[width=0.95\columnwidth]{Figures/AoI_vs_p_V4.eps}}
%
%  \caption{ The impact of channel access probability on the peak and average AoI. }
%  \label{fig:AoI_vs_p}
%\end{figure}

\begin{figure*}[t!]
  \centering

  \subfigure[\label{fig:1a}]{\includegraphics[width=0.95\columnwidth]{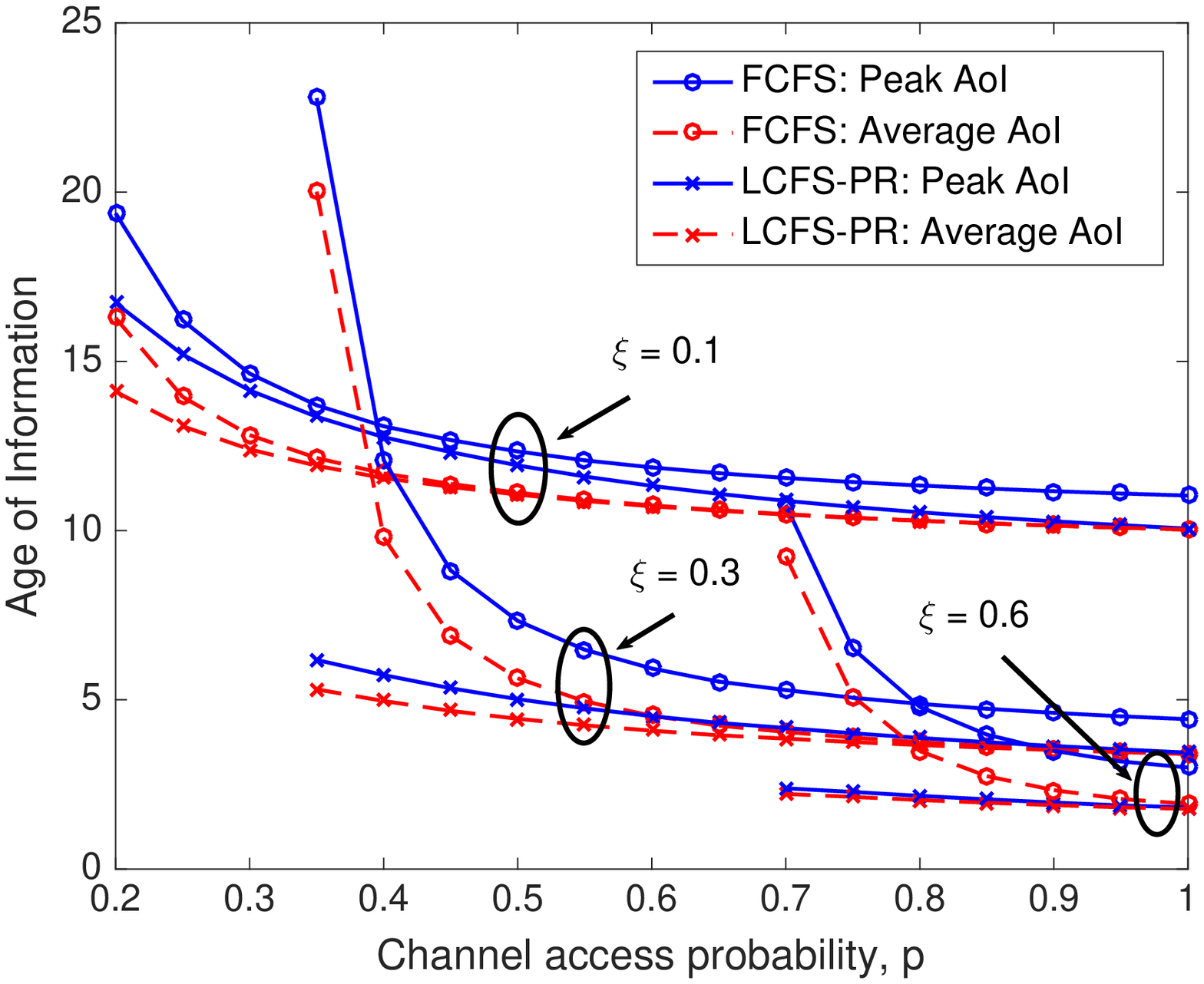}} ~
  \subfigure[\label{fig:1b}]{\includegraphics[width=0.95\columnwidth]{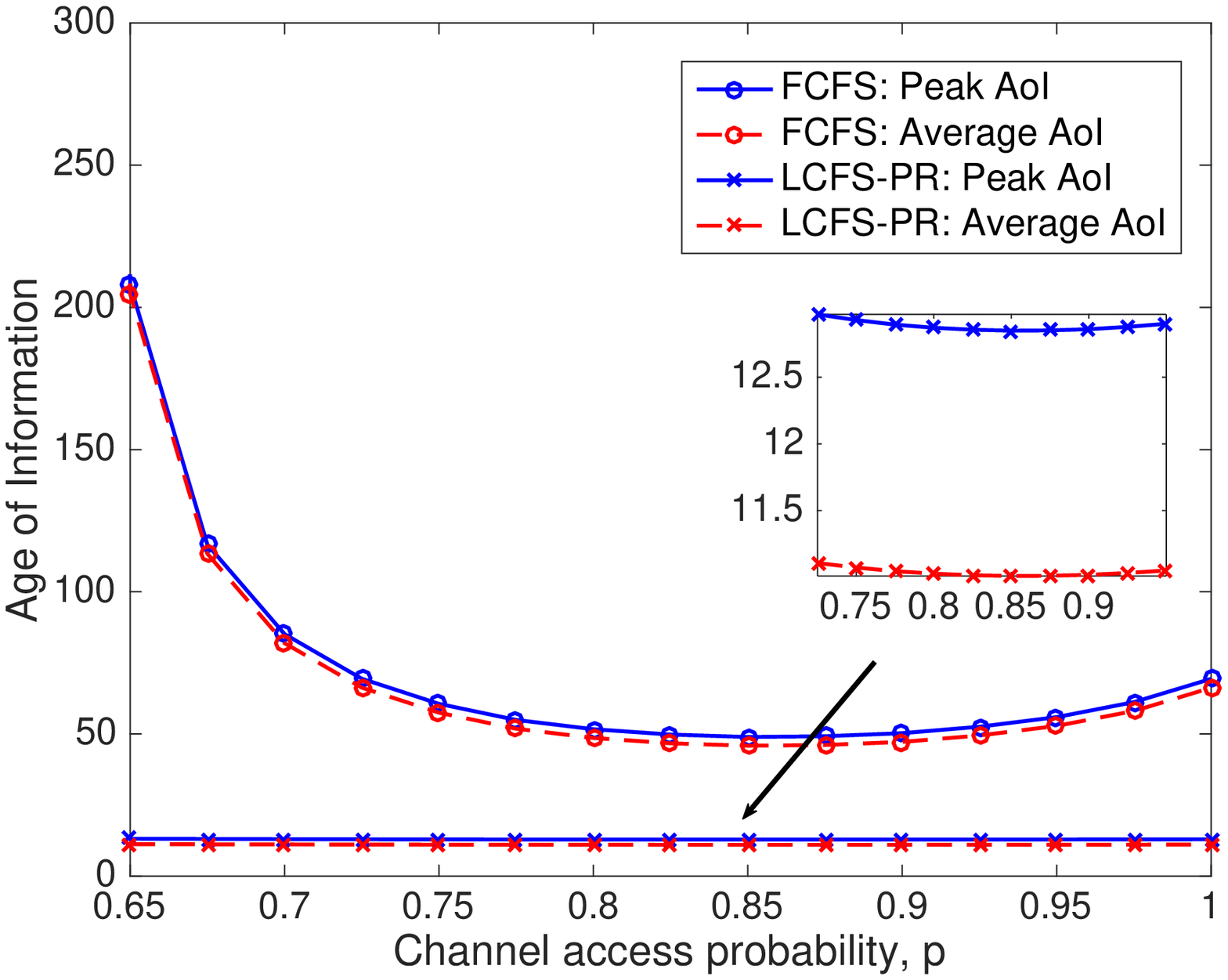}}
  \caption{ The impact of channel access probability on the peak and average AoI. In Fig. (a), the network has a spatial density as $\lambda = 10^{-4}$~m$^{-2}$ and the packet update frequency varies as $\xi = 0.1, 0.3, 0.6$~dB. In Fig. (b), the network has a spatial density as $\lambda = 2 \times 10^{-3}$~m$^{-2}$ and the packet update frequency is maintained at $\xi = 0.30$. }
  \label{fig:AoI_vs_p}
\end{figure*}

Fig.~\ref{fig:AoI_vs_p} plots the AoI as a function of the channel access probability, under various values of packet update frequencies and spatial deployment densities.
In this figure, the AoI statistics under different network parameters are depicted only within their feasible, namely, the queue stable, regions.
Particularly, Fig.~\ref{fig:AoI_vs_p} (a) shows the AoI statistics of a sparsely deployed network, and from this figure several messages are due: ($i$) at $p = 0.7$, the peak and average AoI under the FCFS protocol first decreases and then increases as $\xi$ grows while the same statistic keeps declining with the packet update frequency, which coincides with the observations drawn in Fig.~6;
($ii$) the peak and average AoI under LCFS-PR protocol are converging at $p=1$, because under the employed set of parameters the typical link attains a relatively high transmission success probability, i.e., $p_{\mathrm{s}} \approx 1$, at $p=1$, which implies such a convergence according to Corollary~3;
and
($iii$) under both FCFS and LCFS-PR disciplines, the peak and average AoI statistics decrease monotonically with respect to the channel access probability $p$, and at $p=1$, the AoI can be further reduced by increasing the packet update frequency.
The last observation is in line with the conclusion drawn from the special case study in Section III-D-2). And it mainly ascribes to the fact that when $p$ decreases, while each transmitter reduces its channel access frequency and hence collectively brings down the overall interference level, which leads to a higher transmission success probability per node, that also results in a reduced frequency of channel access, which in fact deteriorate the link throughput and give rise to a higher AoI.
In this respect, simply adopting ALOHA with a universally fixed channel access probability in sparse networks does not benefit the AoI.
Therefore, one shall opt for more advanced channel access controls, e.g., the ALOHA with locally adaptive channel access probability \cite{YanAraQue:19Globecom}, to boost up the AoI performance in large-scale networks.{\footnote{A thorough discussion about the design of locally adaptive ALOHA scheme and the performance of peak AoI under FCFS can be seen in \cite{YanArafaQue:19}. It is worthwhile to note that using the same design mechanism, and by leveraging the AoI expressions given in Lemma~1 and Lemma~2 of this paper, one can show that the locally adaptive channel access probability given in \cite{YanArafaQue:19} is effective in minimizing the peak and average AoI under both FCFS and LCFS-PR.}}
On the other hand, Fig.~\ref{fig:AoI_vs_p} (b) depicts the change of AoI performance when the infrastructure is densely deployed. From this figure, we can clearly see an optimal channel access probability that minimizes the peak and average AoI under both FCFS and LCFS-PR protocols. Because when there is an abundant number of wireless links in the network, the excessive interference can devastate the throughput of each link. In this context, it is worthwhile to reduce the channel access probability so as to strike a balance between the radio channel utilization per node and the overall interference level, which, as illustrated in the special case studies of Section III-D, attains the minimum AoI by achieving a maximum throughput.

\begin{figure}[t!]
  \centering{}

    {\includegraphics[width=0.95\columnwidth]{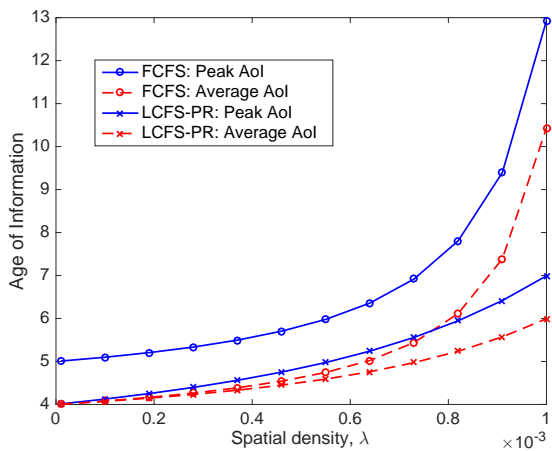}}

  \caption{ The impact of deployment density on the peak and average AoI. }
  \label{fig:AoI_vs_lambda}
\end{figure}
In Fig.~\ref{fig:AoI_vs_lambda}, we put the spotlight on the AoI achieved by a typical UE under the presence of an increase of network deployment density.
From this figure, we find that LCFS-PR always outperforms FCFS in delivering fresh information when the network grows in size.
Additionally, Fig.~\ref{fig:AoI_vs_lambda} also conveys another fundamental message: LCFS-PR is much more suitable than FCFS in densely deployed networks.
In fact, as the deployment density of transceiver pairs goes up from $10^{-5} /\mathrm{m}^{2}$ to $10^{-3}/\mathrm{m}^{2}$, the average AoI under LCFS-PR increases by only 50\% while that under the FCFS more than double, which firmly demonstrates the effectiveness of LCFS-PR.

% ============================================ %
%         Section: Conclusion                  %
% ============================================ %
\section{Conclusion}
In this work, we undertook an analytical study for the understanding of AoI performance in large-scale wireless networks. We used a general model that accounts for the channel gain and interference, dynamics of status updating, and spatially queueing interactions. Our results revealed that, when considering transmission protocols, LCFS-PR is a better resort than FCFS toward minimizing AoI, whereas the gain is more pronounced when the network is densely deployed. Moreover, in a sparsely deployed network, simple channel access controls such as ALOHA cannot reduce the AoI, as long as the access probability is universally devised, and hence ushering a call for more advanced designs. However, when the infrastructure is densely rolled out, there exists a non-trivial ALOHA channel access probability that minimizes the peak and average AoI under both FCFS and LCFS-PR.

The framework provided in this paper allows one to explicitly characterize the effect of spatially interacting queues, which is a fundamental conundrum in analyzing queueing networks. More generally, our work helps to understand how all the key features of a wireless network, i.e., interference, load, and deployment strategy, affect the AoI performance.
In consequence, it opens the door for a variety of further design options, e.g., to explore the impact of different buffer sizes, retransmission schemes, or channel access approaches on the AoI of a large-scale wireless network. Investigating up to what extent power control can improve AoI is also regarded as a concrete direction for future work.

%%%%%%%%%%%%%%%%%%%%%%%%%%%%%%%%%%%%%%%%%%%%%%%%%%%%
\begin{appendix}
\subsection{Proof of Lemma~\ref{lma:Cnd_AoI_LCFS} } \label{apx:Cnd_AoI_LCFS}
Let us denote by $M$ and $N$ the inter-arrival time and the total sojourn time, i.e., the time from a packet's arrival to the time it reaches destination, in the queue, respectively, which are random variables. When conditioning on the point process $\Phi$, the dynamics over the typical link can be regarded as a Geo/Geo/1 queue with service rate being $p \mu^\Phi_0$.
As such, under the LCFS-PR discipline, the average AoI is given as \cite{TriTalMod:19}:
\begin{align} \label{equ:Gnl_AoI_L}
\mathbb{E}^0\big[ A^{ \mathrm{ave} }_{\mathrm{L}} \vert \Phi \big] = \frac{1}{2} \cdot \frac{ \mathbb{E}\big[ M^2 \big] }{ \mathbb{E}[ M ] } + \frac{\mathbb{E} \big[ \min(N, M) \big] }{ \mathbb{P}\big( N \leq M \big) } - \frac{1}{2}.
\end{align}
On the one hand, as $M \sim Geo(\xi)$ and $N \sim Geo( p \mu^\Phi_0)$, we have the following
\begin{align} \label{equ:Mmnt_M}
&\mathbb{E}[M] = \frac{1}{\xi}, \quad
\mathbb{E}[M^2] = \frac{ 2 - \xi }{ \xi^2 }, \\ \label{equ:Prob_NM}
& \mathbb{P}( N \leq M ) = 1 - \mathbb{E}\big[ ( 1 - p \mu^\Phi_0 )^M \big]
\nonumber\\
&\qquad \qquad ~~ = \frac{ p \mu^\Phi_0 }{ 1 - ( 1 - p \mu^\Phi_0 ) ( 1 - \xi ) }.
\end{align}
On the other hand, since $M$ and $N$ are independent random variables, via simple calculations we have $\min(M,N) \sim Geo( 1 - (1-p\mu^\Phi_0)(1-\xi))$. Thus the following holds
\begin{align} \label{equ:Mmnt_min}
\mathbb{E}\big[ \min( M,N ) \big] = \frac{1}{ 1 - ( 1 - p \mu^\Phi_0 )(1-\xi) }.
\end{align}
The result in \eqref{equ:Cnd_AoI_LCFS} then follows from substituting \eqref{equ:Mmnt_M}, \eqref{equ:Prob_NM}, and \eqref{equ:Mmnt_min} into \eqref{equ:Gnl_AoI_L}.

Next, the conditional peak AoI under LCFS-PR is given as \cite{TriTalMod:19}:
\begin{align} \label{equ:Gnl_PAoI_L}
\mathbb{E}^0\big[ A^{ \mathrm{p} }_{\mathrm{L}} \vert \Phi \big] = \frac{ \mathbb{E} [ M ] }{ P\big( N \!\leq\! M \big) } + \frac{ \mathbb{E} \big[ N \mathbbm{1}\{ N \!\leq\! M \} \big] }{ P\big( N \!\leq\! M \big) } - 1.
\end{align}
The nominator of the second term on the R.H.S. above can be calculated as
\begin{align} \label{equ:Epct_NleqM}
 & \mathbb{E} \big[ N \mathbbm{1}\{ N \leq M \} \big] = \mathbb{E}\Big[ \mathbb{E} \big[ N \mathbbm{1}\{ N \leq M \} \big\vert M \big]  \Big]
\nonumber\\
= & \, \mathbb{E}\Big[ \sum_{m=1}^{M} m ( 1 - p \mu^\Phi_0 )^{m-1} p \mu^\Phi_0  \Big]
\nonumber\\
= & \, \mathbb{E} \Big[ \, \frac{ 1 - (1 - p \mu^\Phi_0 )^M }{\mu^\Phi_0} - (1-p \mu^\Phi_0) M ( 1 - p \mu^\Phi_0 )^{M-1} \Big]
\nonumber\\
= & \, \frac{ p \mu^\Phi_0  }{ \big[ 1 - ( 1 - p \mu^\Phi_0 )( 1 - \xi ) \big]^2 }.
\end{align}
The expression in \eqref{equ:Cnd_PAoI_LCFS} then follows by substituting \eqref{equ:Mmnt_M}, \eqref{equ:Prob_NM}, and \eqref{equ:Epct_NleqM} into \eqref{equ:Gnl_PAoI_L}.

\subsection{Proof of Theorem~\ref{thm:Meta_SINR}} \label{apx:Meta_SINR}
For ease of exposition, let us denote $q_{u,j} = \mathbb{P}( \zeta_j = 1 | \Vert y_j - y_0 \Vert = u, \zeta_0 = 1 )$, and $Y_0^\Phi = \ln \mathbb{P}(\gamma_0 > \theta | \Phi)$. Given the typical link is transmitting, by using Lemma~\ref{lma:Cnd_TXS_Prob}, we can write the moment generating function of $Y^\Phi_0$ as follows:
\begin{align} \label{equ:MG_MY0}
& \mathcal{M}_{Y^\Phi_0}(s) = \mathbb{E}\big[ \mathbb{P}(\gamma_0 > \theta | \Phi )^s \big]
\nonumber\\
&= e^{ - \frac{ \theta s r^\alpha }{ \rho } } \mathbb{E}^0 \Big[ \prod_{j \neq 0} \! \big( 1 - \frac{ p \, q_{u,j} }{ 1 + \Vert X_j - y_0 \Vert^\alpha / \theta r^\alpha } \big) \Big]
\nonumber\\
& \stackrel{(a)}{=} e^{ - \frac{ \theta s r^\alpha }{ \rho } } \mathbb{E}^0_{ \hat{\Phi} }\Big[ \prod_{j \neq 0} \! \big( 1 - \frac{ p \, q_{u,j} }{ 1 \!+\! \vert u^2 \!+\! r^2 \!-\! 2 u r \cos \Psi \vert^{\frac{\alpha}{2}} \! / \theta r^\alpha } \big)^s \Big]
\nonumber\\
& \stackrel{(b)}{=} \exp\!\bigg\{ - \frac{ \theta s r^\alpha }{ \rho } \! -  \lambda \! \int_{0}^{\infty} \!\!   \int_{0}^{2\pi} \! d \varphi u du
\nonumber\\
&\quad \times \Big( 1 - \mathbb{E} \big[ \big( 1 - \frac{ p \cdot q_{u,j} }{ 1 \!+\! \vert u^2 \!+\! r^2 \!-\! 2 u r \cos \varphi \vert^{\frac{\alpha}{2}} \! / \theta r^\alpha } \big)^s \big] \Big) \bigg\}
\nonumber\\
& \stackrel{(c)}{=} \exp\!\bigg\{ \!\! -\! \frac{ \theta s r^\alpha }{ \rho } \! - \! \! \int_{0}^{\infty} \!\! \!\!  \int_{0}^{2\pi} \!\sum_{k=1}^{s} \!\! \binom{s}{k}
\nonumber\\
& \qquad \qquad \qquad \qquad \times \frac{ \lambda \, (-p)^{k+1} \, \mathbb{E}[ q^k_{u,j} ] d \varphi u du }{ \big[ 1 \!+\! \vert u^2 \!+\! r^2 \!-\! 2 u r \cos \varphi \vert^{\frac{\alpha}{2}} \! / \theta r^\alpha \big]^k }  \bigg\}
\nonumber\\
& \stackrel{(d)}{=} \exp\!\bigg\{ \!\! -\! \frac{ \theta s r^\alpha }{ \rho } \! - \! 2 \pi \lambda \int_{0}^{\infty} \!\!  \int_{0}^{2\pi}  \sum_{k=1}^{s}   \binom{s}{k}
\nonumber\\
& \qquad \qquad \times {  (-1)^{k+1} \, \mathbb{E}[ q^k_{u,j} ] \, \Big[ 1 - \frac{1}{ p \mathcal{H}_\theta( \frac{u}{r}, \varphi, p ) } \Big]^k  \frac{d \varphi}{ 2 \pi } u du }   \bigg\}
\end{align}
where ($a$) is to take the expectation of point process $\Phi$ by conditioning on the locations of receivers $\hat{\Phi}$, in which a typical pair is depicted per Fig.~\ref{fig:NodeLoc_v1}, and using the cosine law, ($b$) follows from the probability generating functional (PGFL) of a PPP \cite{BacBla:09}, ($c$) aims to further expand the expression via the Binomial theorem, and ($d$) follows from substituting \eqref{equ:H_theta} into the equation and algebraic manipulation.
In order to obtain a complete expression of \eqref{equ:MG_MY0}, we need to further compute $\mathbb{E}[q^k_{u,j}]$.
By using Lemma~\ref{lam:actv_prob}, we arrive at the following:
\begin{align} \label{equ:q_k}
& \mathbb{E}\big[ q^k_{u,j} \big] = \mathbb{E}\big[ \mathbbm{1}\{ p \cdot \mu_j^\Phi \leq \xi \} \vert \Vert y_j - y_0 \Vert = u, \zeta_0 = 1 \big]
\nonumber\\
&+ \mathbb{E} \big[ (\xi/p \mu_j^\Phi)^k \mathbbm{1}\{ p \cdot \mu_j^\Phi > \xi \} \vert \Vert y_j - y_0 \Vert = u, \zeta_0 = 1 \big].
\end{align}
At this stage, let us assume the CDF of $\mu^\Phi_0$, $F_\theta(u)$, is available. The first term on the right hand side (R.H.S.) of \eqref{equ:q_k} can then be computed as follows:
\begin{align} \label{equ:Term1_RHS}
& \mathbb{E}\big[ \mathbbm{1}\{ p \cdot \mu_j^\Phi \leq \xi \} \vert \Vert y_j - y_0 \Vert = u, \zeta_0 = 1 \big]
\nonumber\\
= & \, \mathbb{P} \Big( \mu^{\Phi^{!o}}_j \!\!< \frac{\xi}{p} \cdot \frac{ 1 + D_{0j} }{ 1 + D_{0j} - p } \big\vert \Vert y_j - y_0 \Vert = u, \zeta_0 = 1 \Big)
\nonumber\\
\stackrel{(a)}{=} & \!\! \int_{0}^{2\pi} \!\!\!\!\!\! F_\theta \Big(\, \frac{\xi}{p} \!+\! \frac{\xi}{ 1 \!-\! p \!+\! ( r^2 \!+\! u^2 \!-\! 2 u r \cos \varphi )^{\frac{\alpha}{2}} \!/\theta r^\alpha } \Big) \frac{ d \varphi }{ 2 \pi }
\nonumber\\
= & \!\! \int_{0}^{2\pi} \!\!\!\!\!\! F_\theta \Big(\, \mathcal{H}_{\theta}\big( \frac{u}{r}, \varphi, p \big) \Big) \frac{ d \varphi }{ 2 \pi },
\end{align}
where $\mu^{\Phi^{!o}}_j$ denotes a reduced point process by removing the points associated with the typical link from $\Phi$, and ($a$) follows by using the Slivnyark's theorem \cite{BacBla:09}.
Analogously, we can obtain the expression for the second term on the R.H.S. of \eqref{equ:q_k} as follows:
\begin{align} \label{equ:Term2_RHS}
& \mathbb{E} \Big[ \big( \frac{\xi}{ p \mu_j^\Phi} \big)^k \mathbbm{1}\{ p \cdot \mu_j^\Phi > \xi \} \vert \Vert y_j - y_0 \Vert = u, \zeta_0 = 1 \Big]
\nonumber\\
= & \int_{0}^{2\pi} \!\!\!\! \int_{ \mathcal{H}_{\theta}(\frac{u}{r}, \psi, p) }^{1} \!\!\!\! t^{-k} \mathcal{H}^k_{\theta}(\frac{u}{r}, \psi, p) F_\theta(dt) \frac{d \psi}{ 2 \pi }.
\end{align}

By using the Gil-Pelaze theorem \cite{Gil}, we have the CDF of $\mu^\Phi_0$ given as follows:
\begin{align}
F_\theta(u) &= \mathbb{P}\big[ \mathbb{P}( \gamma_0 > \theta \vert \Phi ) < u \big] = \mathbb{P}( Y^\Phi_0 < \ln u )
\nonumber\\
&= \frac{1}{2} - \frac{1}{\pi} \int_{0}^{\infty} \mathrm{Im} \big\{ u^{-j\omega} \mathcal{M}_{ Y^\Phi_0 }(j \omega) \big\} \frac{d \omega }{ \omega }.
\end{align}
The expression in \eqref{equ:Meta_Grl} then follows by plugging back \eqref{equ:MG_MY0}, \eqref{equ:Term1_RHS}, and \eqref{equ:Term2_RHS} into the above equation and further simplify the expression via \textit{change of variable} as $v = u / r$.

Finally, since $F_\theta(u)$ is given in the form of a fix-point equation, we can iteratively solve for the exact expression via similar approach as \cite{YanQue:19}\footnote{ Such an approach to solve the fixed-point equation via successive approximations is known as the Picard's method \cite{Pic:1893}. }.

\begin{figure}[t!]
  \centering{}

    {\includegraphics[width=0.90\columnwidth]{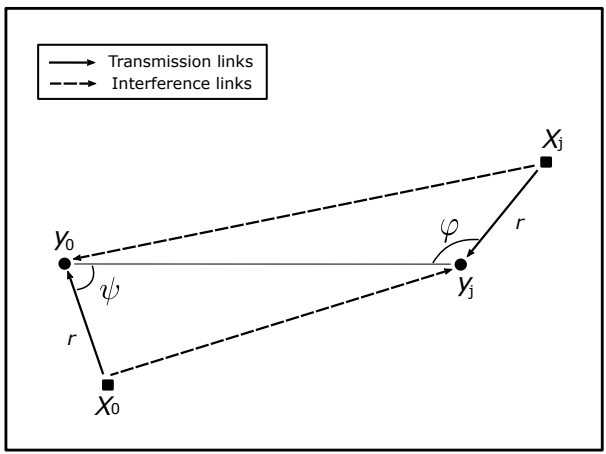}}

  \caption{ Example of a two-points location topology given the distance between receivers being $\Vert y_0 - y_j\Vert = u$. }
  \label{fig:NodeLoc_v1}
\end{figure}

\subsection{Proof of Theorem~\ref{thm:stability}} \label{apx:stability}
Because the dynamics at the typical link can be abstracted as a queueing system with uniform arrivals and a random departure rate, then, according to Loynes' theorem \cite{Loy:62}, the transmissions over the typical link are stable if and only if the packet arrival rate does not exceed the mean departure rate, i.e.,
\begin{align}
\xi \leq p \cdot \mathbb{P}(\gamma_{0,t} > \theta ) = p \cdot p_{ \mathrm{s} }.
\end{align}
The R.H.S. of the above inequality is the mean of the random quantity given by (6), which can be derived via two steps. First, let us assume the transmission success probability $p_{ \mathrm{s} }$ is currently available. Then, given the typical node is transmitting data, the conditional transmission success probability at a generic link $j$ per (11) can be written as follows:
\begin{align}\label{equ:mu_j_cnd}
\mu^\Phi_j &= e^{ - \frac{ \theta r^\alpha }{ \rho } } \prod_{ i \neq 0, i \neq j } \Big( 1 - \frac{ p a^\Phi_i }{ 1 + D_{ij} } \Big) \times \Big( 1 - \frac{ 1 }{ 1 + D_{0j} } \Big)
\nonumber\\
&= \mu^{\Phi^{! 0}}_j \times \Big( 1 - \frac{ 1 }{ 1 + D_{0j} } \Big)
\end{align}
where $\mu^{\Phi^{! 0}}_j$ is the conditional transmission success probability of link $j$ given the reduced Point process $\Phi^{!0}$, which is obtained by removing the dipoles of link 0 from the the original point process $\Phi$.
Using the expression in \eqref{equ:mu_j_cnd}, and by denoting the distance between two receivers as $\Vert y_0 - y_j \Vert = u$, we can calculate the active probability at node $j$ as
\begin{align}
&\mathbb{P}( \zeta_j = 1 \vert \Vert y_0 - y_j \Vert = u, \zeta_0 = 1 )
\nonumber\\
&= \mathbb{E} \Big[ \min\big\{ \frac{ \xi }{ p \mu^\Phi_j }, 1 \big\} \big\vert \Vert y_0 - y_j \Vert = u, \zeta_0 = 1 \Big]
\nonumber\\
&\stackrel{(a)}{=}\! \mathbb{E} \Big[\! \min\!\big\{ \frac{ \xi }{ p \mu^{\Phi^{!0}}_j } \Big( 1 + \frac{ 1 }{ D_{0j} } \Big) , 1 \big\} \big\vert \Vert y_0 - y_j \Vert = u, \zeta_0 = 1 \Big]
\nonumber\\
&\stackrel{(b)}{\approx} \! \mathbb{E} \Big[\! \min\!\big\{ \frac{ \xi }{ p \mathbb{E}[ \mu^{\Phi^{!0}}_j ] } \Big( 1 \!+\! \frac{ 1 }{ | u^2 \!+\! r^2 \!-\! 2 u r \cos \Psi |^{ \alpha / 2 } \!/ \theta r^\alpha } \Big) , 1 \big\} \Big]
\nonumber\\
&\stackrel{(c)}{=} \! \int_{0}^{2\pi} \!\!\!\!\!\! \min\!\Big\{ \frac{\xi}{ p p_{\mathrm s} } \big[ 1 + \frac{ \theta r^\alpha }{ ( u^2 + r^2 - 2 u r \cos \psi )^{ \alpha / 2 } } \big], 1 \Big\} \frac{d \psi}{ 2 \pi }
\end{align}
where ($a$) is by substituting (17), ($b$) by taking expectation directly to the term $\mu^{\Phi^{!0}}_j$ in the denominator, as well as leveraging the Cosine Rule to expand $D_{0j}$, and ($c$) follows from using the Slivnyark's theorem, which gives $\mathbb{E}[\mu^{\Phi^{!0}}_j] = p_{\mathrm{s}}$.

As such, in the steady state, we can approximate the locations of the interfering nodes using a non-homogeneous PPP with spatial density:
\begin{align}
G(u) = \mathbb{P}( \zeta_j = 1 \vert \Vert y_0 - y_j \Vert = u, \zeta_0 = 1 ).
\end{align}
To this end, we can take an expectation on both sides of \eqref{equ:CndTX_Prob} and get the following:
\begin{align}
p_{ \mathrm{s} } \!&=\! e^{ - \frac{\theta r^\alpha}{\rho} } \mathbb{E}\Big[ \prod_{ j \neq 0 } \big( 1 - \frac{ p a^\Phi_j }{ 1 + D_{j0} } \big) \Big]
\nonumber\\
&=\! e^{ - \frac{\theta r^\alpha}{\rho} } \mathbb{E}^0_{\hat{\Phi}}\Big[ \prod_{ j \neq 0 } \mathbb{E}\big[ 1 - \frac{ p a^\Phi_j }{ 1 \!+\! D_{j0} } \big \vert \Vert y_0 - y_j \Vert \!=\! u \big]  \Big]
\nonumber\\
&=\! \exp\!\Big(\!\! -\! \frac{ \theta r^\alpha }{ \rho } \!- \!\! \int_{0}^{\infty} \!\!\!\!\! \int_{0}^{2\pi} \!\!\!\!\! \frac{ \lambda \, p\, \theta\, r^\alpha G(u) u du d \varphi }{  \theta r^\alpha \!+\! \big( u^2 \!+\! r^2 \!-\! 2 u r \cos \varphi \big)^{ \frac{\alpha}{2} }  } \Big).
\end{align}
The result then follows from algebraic manipulations.

\end{appendix}

%\balance
\bibliographystyle{IEEEtran}
\bibliography{bib/StringDefinitions,bib/IEEEabrv,bib/howard_PBN_AoI}

\begin{thebibliography}{10}
\providecommand{\url}[1]{#1}
\csname url@samestyle\endcsname
\providecommand{\newblock}{\relax}
\providecommand{\bibinfo}[2]{#2}
\providecommand{\BIBentrySTDinterwordspacing}{\spaceskip=0pt\relax}
\providecommand{\BIBentryALTinterwordstretchfactor}{4}
\providecommand{\BIBentryALTinterwordspacing}{\spaceskip=\fontdimen2\font plus
\BIBentryALTinterwordstretchfactor\fontdimen3\font minus
  \fontdimen4\font\relax}
\providecommand{\BIBforeignlanguage}[2]{{%
\expandafter\ifx\csname l@#1\endcsname\relax
\typeout{** WARNING: IEEEtran.bst: No hyphenation pattern has been}%
\typeout{** loaded for the language `#1'. Using the pattern for}%
\typeout{** the default language instead.}%
\else
\language=\csname l@#1\endcsname
\fi
#2}}
\providecommand{\BIBdecl}{\relax}
\BIBdecl

\bibitem{KauYatGru:12}
S.~Kaul, R.~Yates, and M.~Gruteser, ``Real-time status: {How} often should one
  update?'' in \emph{Proc. IEEE INFOCOM}, Orlando, FL, Mar. 2012, pp.
  2731--2735.

\bibitem{KosPapAng:17}
A.~Kosta, N.~Pappas, and V.~Angelakis, ``Age of information: {A} new concept,
  metric, and tool,'' \emph{Foundations and Trends in Networking}, vol.~12,
  no.~3, pp. 162--259, 2017.

\bibitem{KamKomEph:14}
C.~Kam, S.~Kompella, and A.~Ephremides, ``Effect of message transmission
  diversity on status age,'' in \emph{Proc. IEEE Int. Symp. Inform. Theory},
  Honolulu, HI, Jun. 2014, pp. 2411--2415.

\bibitem{CosCodEph:16}
M.~Costa, M.~Codreanu, and A.~Ephremides, ``On the age of information in status
  update systems with packet management,'' \emph{IEEE Trans. Inf. Theory},
  vol.~62, no.~4, pp. 1897--1910, Feb. 2016.

\bibitem{HuaMod:15}
L.~Huang and E.~Modiano, ``Optimizing age-of-information in a multi-class
  queueing system,'' in \emph{Proc. IEEE Int. Symp. Inform. Theory}, Hong Kong,
  China, Jun. 2015, pp. 1681--1685.

\bibitem{Yat:18}
R.~D. Yates, ``Status updates through networks of parallel servers,'' in
  \emph{Proc. IEEE Int. Symp. Inform. Theory}, Vail, CO, Jun. 2018, pp.
  2281--2285.

\bibitem{YatKau:18}
R.~D. Yates and S.~K. Kaul, ``The age of information: {Real-time} status
  updating by multiple sources,'' \emph{IEEE Trans. Inf. Theory}, vol.~65,
  no.~3, pp. 1807--1827, Sept. 2018.

\bibitem{KadUysSin:16}
I.~Kadota, E.~Uysal-Biyikoglu, R.~Singh, and E.~Modiano, ``Minimizing the age
  of information in broadcast wireless networks,'' in \emph{Proc. IEEE
  Allerton}, Monticello, IL, Sept. 2016, pp. 844--851.

\bibitem{KadSinUys:18}
I.~Kadota, A.~Sinha, E.~Uysal-Biyikoglu, R.~Singh, and E.~Modiano, ``Scheduling
  policies for minimizing age of information in broadcast wireless networks,''
  \emph{IEEE/ACM Trans. Netw.}, vol.~26, no.~6, pp. 2637--2650, Dec. 2018.

\bibitem{TalKarMod:18}
R.~Talak, S.~Karaman, and E.~Modiano, ``Optimizing age of information in
  wireless networks with perfect channel state information,'' in \emph{Proc.
  Modeling and Optimization in Mobile, Ad Hoc, and Wireless Networks (WiOpt)},
  Shanghai, China, May 2018, pp. 1--8.

\bibitem{YanAraQue:20ICASSP}
H.~H. Yang, A.~Arafa, T.~Q.~S. Quek, and H.~V. Poor, ``Age-based scheduling
  policy for federated learning in mobile edge networks,'' in \emph{Proc. IEEE
  Int. Conf. Acoustics, Speech, and Signal Processing}, 2020, available as
  ArXiv: 1910.14648.

\bibitem{DevDurFer:19}
R.~Devassy, G.~Durisi, G.~C. Ferrante, O.~Simeone, and E.~Uysal, ``Reliable
  transmission of short packets through queues and noisy channels under latency
  and peak-age violation guarantees,'' \emph{IEEE J. Sel. Areas Commun.},
  vol.~37, no.~4, pp. 721--734, Apr. 2019.

\bibitem{HeYuaEph:16}
Q.~He, D.~Yuan, and A.~Ephremides, ``Optimizing freshness of information: {On}
  minimum age link scheduling in wireless systems,'' in \emph{Proc. Modeling
  and Optimization in Mobile, Ad Hoc, and Wireless Networks (WiOpt)}, Tempe,
  AZ, May 2016, pp. 1--8.

\bibitem{talak2018optimizing}
R.~Talak, S.~Karaman, and E.~Modiano, ``Optimizing information freshness in
  wireless networks under general interference constraints,'' \emph{arXiv
  preprint arXiv:1803.06467}, 2018.

\bibitem{XuYanWan:19}
C.~Xu, H.~H. Yang, X.~Wang, and T.~Q.~S. Quek, ``Optimizing information
  freshness in computing enabled {IoT} networks,'' \emph{{IEEE} Internet of
  Things Journal}, 2019.

\bibitem{CorRohGun:19}
L.~Corneo, C.~Rohner, and P.~Gunningberg, ``Age of information-aware scheduling
  for timely and scalable internet of things applications,'' in \emph{Proc.
  IEEE INFOCOM}, Paris, France, Jun. 2019, pp. 2476--2484.

\bibitem{YanGerZho:17}
H.~H. Yang, G.~Geraci, Y.~Zhong, and T.~Q.~S. Quek, ``Packet throughput
  analysis of static and dynamic {TDD} in small cell networks,'' \emph{IEEE
  Wireless Commun. Lett.}, vol.~6, no.~6, pp. 742--745, Dec. 2017.

\bibitem{WanYanZhu:19WCL}
Y.~Wang, H.~H. Yang, Q.~Zhu, and T.~Q.~S. Quek, ``Analysis of packet throughput
  in spatiotemporal hetnets with scheduling and various traffic loads,''
  \emph{IEEE Wireless Commun. Lett.}, 2019.

\bibitem{ZhoQueGe:16}
Y.~Zhong, T.~Q.~S. Quek, and X.~Ge, ``Heterogeneous cellular networks with
  spatio-temporal traffic: {Delay} analysis and scheduling,'' \emph{IEEE J.
  Sel. Areas Commun.}, vol.~35, no.~6, pp. 1373--1386, Jun. 2017.

\bibitem{GhaElsBad:17}
M.~Gharbieh, H.~ElSawy, A.~Bader, and M.-S. Alouini, ``Spatiotemporal
  stochastic modeling of {IoT} enabled cellular networks: Scalability and
  stability analysis,'' \emph{IEEE Trans. Commun.}, vol.~65, no.~9, pp.
  3585--3600, Aug. 2017.

\bibitem{YanWanQue:18}
H.~H. Yang, Y.~Wang, and T.~Q.~S. Quek, ``Delay analysis of random scheduling
  and round robin in small cell networks,'' \emph{IEEE Wireless Commun. Lett.},
  vol.~7, no.~6, pp. 978--981, Dec. 2018.

\bibitem{YanQue:19}
H.~H. Yang and T.~Q.~S. Quek, ``Spatiotemporal analysis for {SINR} coverage in
  small cell networks,'' \emph{IEEE Trans. Commun.}, vol.~67, no.~8, pp. 5520
  -- 5531, May 2019.

\bibitem{ChiElSCon:17}
G.~Chisci, H.~ElSawy, A.~Conti, M.-S. Alouini, and M.~Z. Win, ``On the
  scalability of uncoordinated multiple access for the internet of things,'' in
  \emph{Int. Symposium on Wireless Commun. Systems (ISWCS)}, Bologna, Italy,
  Aug. 2017, pp. 402--407.

\bibitem{ChiElSCon:19}
------, ``Uncoordinated massive wireless networks: Spatiotemporal models and
  multiaccess strategies,'' \emph{{IEEE/ACM} Trans. Networking}, vol.~27,
  no.~3, pp. 918--931, Jun. 2019.

\bibitem{HuZhoZha:18}
Y.~Hu, Y.~Zhong, and W.~Zhang, ``Age of information in poisson networks,'' in
  \emph{Proc. Int. Conf. Wireless Commun. and Signal Process. (WCSP)},
  Hangzhou, China, Dec. 2018, pp. 1--6.

\bibitem{BacBla:09}
F.~Baccelli and B.~Blaszczyszyn, \emph{Stochastic Geometry and Wireless
  Networks. Volumn I: Theory}.\hskip 1em plus 0.5em minus 0.4em\relax Now
  Publishers, 2009.

\bibitem{TriTalMod:19}
V.~Tripathi, R.~Talak, and E.~Modiano, ``Age of information for discrete time
  queues,'' \emph{Available as ArXiv:1901.10463}, 2019.

\bibitem{YanAraQue:20Globecom}
H.~H. Yang, A.~Arafa, T.~Q.~S. Quek, and H.~V. Poor, ``Age of information in
  random access networks: A spatiotemporal study,'' in \emph{Proc. IEEE Global
  Commun. Conf. (Globecom)}, Taipei, Taiwan, Dec. 2020, pp. 1--6.

\bibitem{ManAbdDhi:20}
P.~D. Mankar, M.~A. Abd-Elmagid, and H.~S. Dhillon, ``Spatial distribution of
  the mean peak age of information in wireless networks,'' \emph{Available as
  ArXiv:2006.00290}, 2020.

\bibitem{GupKum:00}
P.~Gupta and P.~R. Kumar, ``The capacity of wireless networks,'' \emph{{IEEE}
  Trans. Inf. Theory}, vol.~46, no.~2, pp. 388--404, Mar. 2000.

\bibitem{haenggi2016meta}
M.~Haenggi, ``The meta distribution of the {SIR} in poisson bipolar and
  cellular networks,'' \emph{IEEE Trans. Wireless Commun.}, vol.~15, no.~4, pp.
  2577--2589, Apr. 2016.

\bibitem{SanBac:17}
A.~Sankararaman and F.~Baccelli, ``Spatial birth--death wireless networks,''
  \emph{IEEE Trans. Inf. Theory}, vol.~63, no.~6, pp. 3964--3982, Jun. 2017.

\bibitem{SanBacFos:19}
A.~Sankararaman, F.~Baccelli, and S.~Foss, ``Interference queueing networks on
  grids,'' \emph{Ann. Applied Probability}, vol.~29, no.~5, pp. 2929--2987,
  2019.

\bibitem{Har:13}
M.~Harchol-Balter, \emph{Performance modeling and design of computer systems:
  queueing theory in action}.\hskip 1em plus 0.5em minus 0.4em\relax Cambridge
  University Press, Cambridge, 2013.

\bibitem{AndAsk:00}
G.~E. Andrews, R.~Askey, and R.~Roy, \emph{Special functions}.\hskip 1em plus
  0.5em minus 0.4em\relax Cambridge University Press, Cambridge, 2000.

\bibitem{ZhaYanShe:20}
X.~Zhang, H.~H. Yang, C.~Shen, G.~Zhu, and T.~Q.~S. Quek, ``{SIR} coverage
  analysis in multi-cell downlink systems with spatially correlated queues,''
  \emph{IEEE Access}, vol.~8, pp. 99\,832 -- 99\,845, apr. 2020.

\bibitem{BacRybShl:18}
F.~Baccelli, A.~Rybko, S.~Shlosman, and A.~Vladimirov, ``Metastability of
  queuing networks with mobile servers,'' \emph{J. Stat. Physics}, vol. 173,
  no. 3-4, pp. 1227--1251, 2018.

\bibitem{WanZhoRee:13}
H.~Wang, X.~Zhou, and M.~C. Reed, ``Physical layer security in cellular
  networks: A stochastic geometry approach,'' \emph{IEEE Trans. Wireless
  Commun.}, vol.~12, no.~6, pp. 2776--2787, Jun. 2013.

\bibitem{ChaRyuQue:15}
S.~H. Chae, J.~Y. Ryu, T.~Q. Quek, and W.~Choi, ``Cooperative transmission via
  caching helpers,'' in \emph{Proc. IEEE Global Commun. Conf. (Globecom)}, San
  Diego, CA, Dec. 2015, pp. 1--6.

\bibitem{YanAraQue:19Globecom}
H.~H. Yang, A.~Arafa, T.~Q.~S. Quek, and H.~V. Poor, ``Locally adaptive
  scheduling policy for optimizing information freshness in wireless
  networks,'' in \emph{Proc. IEEE Global Commun. Conf. (Globecom)}, Waikoloa,
  HI, Dec. 2019, pp. 1--6.

\bibitem{YanArafaQue:19}
------, ``Optimizing information freshness in wireless networks: A stochastic
  geometry approach,'' \emph{IEEE Trans. Mobile Comput.}, 2020.

\bibitem{Gil}
J.~Gil-Pelaez, ``Note on the inversion theorem,'' \emph{Biometrika}, vol.~38,
  no. 3-4, pp. 481--482, Dec. 1951.

\bibitem{Pic:1893}
E.~Picard, ``Sur l'application des m{\'e}thodes d'approximations successives
  {\`a} l'{\'e}tude de certaines {\'e}quations diff{\'e}rentielles
  ordinaires,'' \emph{J. de Math. Pures et Appliqu{\'e}es}, vol.~9, pp.
  217--272, 1893.

\bibitem{Loy:62}
R.~M. Loynes, ``The stability of a queue with non-independent inter-arrival and
  service times,'' in \emph{Math. Proc. Cambridge Philos. Soc.}, vol.~58,
  no.~3.\hskip 1em plus 0.5em minus 0.4em\relax Cambridge University Press,
  1962, pp. 497--520.

\end{thebibliography}

\end{document}